\documentclass[12pt]{revtex4} 
\usepackage[dvips]{graphicx,color}
\def\bea{\begin{eqnarray}}
\def\eea{\end{eqnarray}}
\def\be{\begin{equation}}
\def\ee{\end{equation}}
\def\nn{\nonumber}

\def\d{\delta}

\def\e{\epsilon}

\def\g{\gamma}
\def\G{\Gamma}
\def\k{\kappa}
\def\k{\kappa}
\def\l{\lambda}

\def\m{\mu}
\def\n{\nu}

\def\r{\rho}
\def\s{\sigma}

\def\t{\tau}

\def\a{\alpha}
\def\b{\beta}\def\th{\theta}

\begin{document}
{\flushright{\tiny{UNB-Technical-Report 04/03}}}
\title{Semi-classical quantisation of space-times with apparent horizons.}

\vspace{1cm}

\author{\sc Arundhati Dasgupta}
\affiliation{
  Department of Mathematics and Statistics, University of New Brunswick, Fredericton, Canada E3B 5A3}
\email{dasgupta@math.unb.ca,adasgupt@unb.ca}
\begin{abstract}
Coherent or semiclassical states in canonical quantum gravity describe the classical Schwarzschild space-time. 
By tracing over the coherent state wavefunction inside the horizon, a density matrix is derived. Bekenstein-Hawking entropy is 
 obtained from the density matrix, modulo the Immirzi parameter. The expectation value of the area and curvature operator
is evaluated in these states. The behaviour near the singularity of the curvature operator shows that the
singularity is resolved. We then generalise the results to space-times with spherically symmetric apparent horizons.
\end{abstract}

\maketitle

\section{Introduction}
Classical Black Holes, are observational realities, however the semi classical physics associated
with them remains to be explained.
The black hole horizon is attributed with entropy, temperature, and a non-unitary  
 form of thermal radiation or Hawking radiation\cite{haw}.
 What is the microscopic origin of entropy? How does a quantum mechanical wavefunction
describe the horizon? Is the quantum black hole a pure state in the quantum theory? Why does
the horizon radiate, and what is the end point of evaporation? Despite many plausible explanations,
little has been achieved to demonstrate the complete truth.

Further, classical general relativity predicts destruction of all material falling inside the horizon 
 due to the presence of a central singularity. Do all the material accreting into the black hole
perish at the central singularity even in a quantum mechanical description? 
The answers to these questions requires a full understanding
of quantum geometry, as at distances of the order of Planck length near the singularity,
quantum effects will dominate over the classical prediction of a curvature singularity.

To obtain such a quantum description, one needs a theory of quantum gravity. However, a complete theory of
quantum gravity does not exist, though glimpses of truth have emerged in 
certain regimes. One such regime has been semiclassical gravity, where previously,
quantum fields in curved space-time \cite{haw} were studied. Gravity remained
classical. 
 However, with the development of non-perturbative quantum gravity,
relevant questions where one could `semiclassically quantise' a given space-time have been answered to
a certain extent. 
Semiclassical states have been constructed in canonical quantum gravity and we discuss the
coherent states in this paper. These states have been well known in quantum mechanics
 and are `wave packets' as opposed to exact eigenstates. They provide the
closest approximation to classical physics as uncertainty is minimum
here and the states are peaked in both the momentum and position representations.
Thus expectation value of both momentum and configuration space variables 
are closest to their classical values as measured in these states. In case of canonical gravity, where
time is separated, and the intrinsic metric of constant time slices and
the extrinsic curvature constitute the phase space, these states can be used to build
a entire space-time.
Here a given black hole space-time
with spatial slicings which include the horizon and the central singularity are discussed.
To locate the apparent horizon and the central singularity one has to measure both the intrinsic metric
and the extrinsic curvature of the spatial slice. This `simultaneous' measurement is possible in the coherent states, as they
are peaked in both the phase space variables, and the uncertainty is minimum. This would have been impossible in any 
other semi-classical state, where measuring the extrinsic curvature would have resulted in a complete
uncertainty of the intrinsic metric. 

Coherent states for gravity were constructed in \cite{thiem}
using a formalism due to Hall. In this article, we address coherent
states for black holes, first introduced in \cite{adg1}, and then used to derive
the entropy of the black hole apparent horizon \cite{adg2}. We give a complete derivation
of a density matrix for the apparent horizon here by tracing over the wavefunction inside the horizon. 
The entropy is then obtained by the definition $S=- {\rm Tr}{\rho\ln \rho}$. This gives 
entropy to be proportional to area of the horizon in the first approximation, modulo a constant, which can be
fixed to $1/4$ due the the Immirzi parameter ambiguity in the formulation of the theory. 
The curvature operator or the Kretschmann scalar expectation value is also studied in detail in the states.  
The central singularity in the classical curvature is clearly resolved in the semiclassical expectation value of the
operator, mainly due to the uncertainty which prevents any measurement of area 0. Thus a upper bound exists for the
curvature operator value proportional to the semiclassical parameter $t$ which measures quantum fluctuations around a given classical geometry. As the semiclassical parameter goes to zero,
the singularity re-appears indicating classical physics. 

The area operator is 
also examined here, as this crucially determines the entropy-area law. The classical value
of the area as measured is given by equispaced numbers, and the entropy is actually proportional
to the degeneracy of the area operator. The apparent horizon equation, which introduces correlations
in the coherent state wavefunction does not impose any additional constraints on the area eigenvalues,
and hence the counting yields a different value of the Immirzi parameter as obtained in \cite{meiss}.

In the first
section, we review the coherent states, and in the next section we give an introduction
to the classical phase space for gravity and discuss applications of coherent states to
the same. The canonical variables, the black hole phase space, the corresponding coherent state, and a evaluation of the
expectation value of the curvature operator are discussed next. The apparent
horizon equation is examined in details, and a method of isolating the boundary conditions
to be imposed on the coherent state wavefunction is analysed. The apparent horizon is a difference
equation in the canonical discretised variables, and introduces correlations across the horizon. When the 
wavefunction inside the horizon is traced the density matrix
describing the black hole space-time is described. The entropy and the Immirzi parameter are discussed, and the
article concludes with a discussion and projects for future. The entire formulation here for the derivation of the
entropy can be extended to include space-times with spherically symmetric apparent horizons. 
\section{Coherent States}
The coherent states are constructed to obtain classical physics from quantum mechanics. The origin of these states 
is well known in quantum mechanics for the simple harmonic oscillator, where
the states appear as eigenstates of the annihilation operator $\hat a |z> = z |z>$,
($z$ is a label and represents a point in the complexified classical phase space $x_{\rm cl} - i p_{\rm cl}$, $x_{cl}$ denotes position and $p_{cl}$ are momentum).
In general, according to Klauder \cite{klauder1, klauder2}, coherent states are labelled by
a continuous parameter $z$, and provide a overcomplete basis for the Hilbert space, for a appropriate measure $d \mu(z)$.
\be
\int |z><z| d \mu(z) = 1
\ee
A further restriction on the $z$ to label points in classical phase space for a given system,
fixes the state uniquely. For the Harmonic Oscillator, the Coherent states are also minimum
uncertainty states or $\Delta x \Delta p = \bar h/2$ (x,p being the configuration and momentum variable). Before going to 
the generalisation for gravity, I quote from \cite{klauder2}:
{\it Classical dynamics is quantum dynamics restricted to the only quantum degrees of freedom that may possibly be varied at a macroscopic 
level, namely, the mean position and the mean momentum (or velocity).} This statement follows from the assumption that 
the classical action principle can be derived from a `quantum action principle' if the states of Hilbert space are restricted
to the coherent states.
The quantum action principle is \cite{klauder1}
\be
I_{\rm quantum} = \int \left[<\psi|\imath \frac{d}{d \t} |\psi> -<\psi|H|\psi> \right] d \t
\label{quant}
\ee
$|\psi>$ is a wavefunction in the Hilbert space, $\t$ is the time parameter, and $H$ the Hamiltonian of the
system. The Schrodinger equation results with the variation of $<\psi|$. However, if one takes
the wavefunction to be of the form 
\be
|\psi> = e^{-i q \hat P} e^{i p \hat Q} |0>, \label{abc}
\ee
where $q,p$ are macroscopic position and momentum labels, and $\hat P,\hat Q$ are corresponding
quantum operators,
 then
the above action principle, with the variation of $p,q$ (hence the coherent states) lead to the following equation:
\be
I_{\rm res quantum} = \int [p\dot q - <H> ] d\t
\label{class}
\ee
Clearly, if $<H>$ is the classical Hamiltonian, then one recovers the classical Hamiltonian
equations of motion.
The observation of \cite{klauder2} is that the coherent states comprise a restricted
set, and in all the exactly solvable systems, the states yield the classical system.
In case of gravity, a different set of definitions have been used to generalise the coherent state structure. The basic features however remain the same.\\
1)The states are labelled by points in the classical phase space, \\
2)They provide a resolution of unity.\\
The new definition of the coherent state generalised to gravity is due to Hall, \cite{hall1}. In a coherent state transform
states in the configuration space ($L^2(R)$)are taken to states in the Hilbert space defined on the holomorphic 
sector of the complexified phase space ($ H(C) \cap L^2 (C)$). The kernel of the transformation is a coherent state wavefunction.  The wavefunction is also 
the analytic continuation of the heat kernel of the Laplacian, which after appropriate normalisation corresponds to the kernel that appears in the kernel of the Coherent state transform
\cite{hall1,hall2}. The coherent state obtained thus, coincides with the harmonic oscillator coherent state wavefunction,
which is a eigenstate of the annihilation operator. The coherent state transform can be now defined for Hilbert spaces
for arbitrary gauge compact gauge groups, in particular for SU(2). Canonical gravity in the Ashtekar-Barbero-Immirzi 
variables takes the form of a SU(2) gauge theory with additional constraints due to diffeomorphism invariance.
The generalisation of the coherent states to a diffeomorphism invariance context appears
in \cite{ashthiem}, and a complete study of their properties are in a series of papers in \cite{thiem,tow,sto}.
A very interesting question arises: Is there a similar `Quantum Action' principle as in (\ref{quant}) for
Gravity, and if so do the coherent states restrict the action to the `Classical action' as in (\ref{class})?
 This indeed is a very difficult question, as no one knows a `corresponding' Schrodinger
equation for the quantum gravity states, and the only known `equivalent' equation, the Wheeler- Dewitt, equation is derived
from the classical action, ($\delta S/\delta N=0 $), where $N$ is the lapse. Hence would not
qualify as a ab initio `quantum equation' (The Schrodinger equation is not derived from the classical action  
for quantum mechanics). 
For the Hall coherent states, the SHO coherent states are recovered in case of $L^2(R)$, and Equation (\ref{abc}) is indeed true.
 However, for the coherent states of gravity, which are defined on SU(2) group, what would be the corresponding equations 
for (\ref{quant}) and (\ref{class}).  We discuss this
in the next section, after identifying the phase space for gravity. 
For the sake of clarity, we define the coherent state as \cite{hall1,hall2},
\be
\psi^t(z) = \rho^t(x)_{x\rightarrow z}
\label{eq:coh}
\ee
where, $\rho$ is the heat kernel for the Laplacian on the given configuration space, (for SHO the configuration space is
$R$). $z$ takes values in the complexified phase space, and corresponds to the continuous 
label of states as per the definition in \cite{klauder1}, and the other 
 parameter $t$, is essentially the `semiclassicallity' parameter. This parameter gives the
width or variance of the coherent state around the mean value or peak value in position space.
In case of the simple harmonic oscillator, expectation values of operators correspond to exact classical values, and hence, 
$<\hat P,\hat Q>$= p,q, irrespective of the semiclassicality parameter. For the coherent 
states in gravity, these statements are true only in the limit $t\rightarrow 0$.  
Hence, any expansion of the expectation value of operators in this parameter $t$ is actually a study of 
quantum fluctuations around a given classical geometry. The parameter $t$, as defined in \cite{thiem}
is defined as $l_p^2/a$, where $l_p$ is the Planck length, and $a$ a dimensionful parameter, which
in the case of Schwarzschild black hole can be $r_g^2$ (horizon radius squared).

Continuing the discussion on the coherent states, the generalisation of the above definition in Equation (\ref{eq:coh}), 
leads to the following for the Hilbert space of any arbitrary gauge group $H$, whose complexified elements lie in $G$.
\be
\psi^t(g) = \rho^t(h)_{h\rightarrow g}
\ee
($\rho^t(h)= exp(-t\nabla)\d_{hh'}$ is the heat kernel of the Laplacian on the group manifold.)
with an appropriate normalisation ($h\subset H, g\subset G$). The states are overcomplete with respect 
to a measure $d\mu(g)$ which in the case of SU(2), was shown to be the Liouville measure \cite{hall1, thiem}. 
The Laplacian for SU(2) corresponds to the Casimir operator which has the eigenvalues $j(j+1)$ in the $jth$ irreducible representation, the coherent state can be written as a sum over the irreducible representations, using a theorem due to Peter and Weyl.
\be
\psi^t(g) = \sum_{j} d_j e^{-t j(j+1)/2} \chi_j(gh^{-1})
\ee
($d_j$ is the degeneracy of the irreducible representation with character $\chi_j$).
Since it is this form which is relevant for Canonical gravity, we proceed to find appropriate phase space variables for gravity, 
and then define the coherent state as a function of the phase space variables. 
\section{Classical Phase space for gravity}
We study gravity with the space-time metric $g_{\m\n}$ as the configuration space variable. Due to 
diffeomorphism invariance, the reduced space is really $g_{\m \n}/Diff(M)$.
A separation of time and space in the ADM formulation further fixes  `the configuration space' as the 
intrinsic metric $q_{ab}$, of the constant time slices, the lapse and the shift for propagation along 
the time like directions,
are given by $N, N_a$, where $a=1,2,3$. One can define the momenta conjugate to these variables from the classical action.  As is well known, the Hamiltonian, which is dual to the lapse $N$, is a constraint in gravity due to the absence of the $\dot N$ term in the action.
Thus, only on the constrained surface, the actual variables are
$q_{ab}$ and the canonical conjugate variable $\pi_{ab} = q^{-1/2}(K_{ab} -q_{ab} K)$ , where
$K_{ab}$ is the extrinsic curvature of the slice. A coherent state for these geometrodynamical
variables is yet to be constructed. To use the Hall coherent state, one has to use the
new variables formulation of canonical gravity.
Here the tangent space of each point on the spatial slice is used to define the variables:
\be
A_a^I = \G_a^I - \b K_{ab} ~E^{b I}, \ \ \ \ \ \ \ \b E^{a}_I E^{b}_I = {\rm det q}~ q^{ab}
\ee
with $\b$ being the arbitrary parameter in the theory or the Immirzi parameter, and $I$ runs from 1,2,3 to denote the $SO(3)$ or $SU(2)$ degrees of freedom of the tangent space. $\G_a^I$ is the spin connection,
$E^{b I}$ are densitised triads and $A^I_a$ is the SU(2) gauge connection. 

In these, the action for gravity has the form
\be
I_{G} = \frac{1}{\kappa}\int d^3 x [\dot A^I_a E^a_I - \Lambda^I G_I - N H - N_a H_a] d\t \label{eq:act}
\ee
The above has the form of a Yang-Mill's action, however with additional constraints in the form of the Hamiltonian $H$, and the
diffeomorphism generators $H_a$ ($G^I$ is the usual SU(2) Gauss's constraint, $\Lambda^I$ is the Lagrange multiplier). On the
classical phase space, the constraint equations are $H, H_a=0$ and the canonically conjugate variables 
are $A_a^I, E^a_I$. Clearly no one knows what the action principle for Quantum Gravity is (or the 
Schrodinger Equation) is for quantum gravity. Does the restriction to Coherent states yield the classical 
equations for the canonical variables?
As we know, that finally the quantisation of the action in Equation (\ref{eq:act}) is carried out
in the smeared variables called the holonomy: $h_e(A)$ which are path ordered exponentials of the
gauge connection along one dimensional analytic edges e, and
the corresponding dual momentum $P^I_e$ which are the densitised triads smeared along 2-dimensional surfaces. This is effectively a discretisation
of space, and one defines the basic variables over graphs, and their duals. In \cite{thiem}, Thiemann explored the variables originally
defined by Ashtekar and Lewandowski further, and succeeded in `quantising' these graph dependent variables, and also obtained the appropriate classical limit for these operators.
The variables are:
\be
h_e(A) = {\cal P} \exp(\int_e A) \ \ \ \  P^I_e(A, E) = \frac{1}{a}{\rm Tr}[T^I h_e \int_S h_{\r}*E h_{\r}^{-1} h_e^{-1}]
\label{eq:disc}
\ee
($T^I= -i {\bf\g^I}$ are the generators of SU(2)($\g^I$ are Pauli Matrices), and $a$ is a dimensionfull parameter, usually fixed as a function of the parameters in the classical theory). The $h_{\rho}$ are the
holonomies along edges defined on the 2-surface. The variables satisfy the Poisson Algebra,
\be
\left\{h_e, h_{e'}\right\}=0, \ \ \ \left\{h_e, P_{e'}^I\right\}= \frac{\k}{a}\delta_{e e'} h_{e} \frac{T^I}{2} \ \ \
\left\{P^I_{e'}, P_e^J\right\} = \frac1{a}\e^{IJK} P^K_e \delta_{ee'}
\ee
The complexified element formed from these phase space variables (similar to $x-\imath p$ in R) is an
element of SL(2,C), and is $g_e= e^{-i T^I P^I/2} h_e$. The Coherent state is constructed to be peaked
at the SL(2,C) valued element. The classical 
 action for these discrete variables (for a particular edge) will be (on the Constrained Surface):
\be
S= \frac{a}{\k}\int d\t ~{\rm Tr}[ T^I ~h^{-1}_e \left(\frac {d h_e}{d\t}\right) P_e^I]
\label{eq:kin}
\ee
If we define the `quantum action' principle for gravity in a similar way, except that one confines
oneself to the derivation of the Kinetic term (\ref{eq:kin}), as above, then, it has the form:
\be
S_q= \imath \int d\t <\xi|\frac{d}{d\t}|\xi>
\ee
where $<\xi|$ is a arbitrary state in the Quantum Hilbert Space, then does one recover (\ref{eq:kin}) by confining oneself to the coherent state? Or in other words, what is
\be
S_{\rm resquantum} = \imath \int d\t <\psi^t|\frac{d}{d\t}|\psi^t>
\ee

(where $\psi^t$ is the coherent state for a single edge.)
Strangely enough, though in a rather straightforward calculation (reported in the Appendix(A))) , one recovers (\ref{eq:kin}), in the limit $t\rightarrow 0$. 

So in principle, we have succeeded in deriving a appropriate phase space for gravity, for which
a coherent state can be defined, in a very similar manner to any other quantum mechanical system, and 
they give classical physics. All the previous discussion is about the coherent state peaked at
classical phase space variables of one particular edge. The entire manifold is however charted with a graph, comprised of edges
linked at vertices. A SU(2) Hilbert space is associated with one edge, and the complete description of the 
entire manifold, is a tensor product of Hilbert spaces of all the edges comprising the graph.
The coherent state for the graph will thus be of the form
\be
\Psi= \prod_e \psi_e
\ee
This is a gauge covariant tensor product, of the Coherent state defined on each edge.
 (The gauge transformations act on the holonomy and the corresponding momenta thus:
\be
h_e\rightarrow g(0)h_e g(1)^{-1} , \ \ \ \ \ \ P^I_e \rightarrow g(0)P^I_e g(0)^{-1}
\ee
The $g(0)$ and $g(1)$ are the SU(2) valued group elements acting at the starting point and end point of a edge
respectively.)
The gauge invariant coherent state has intertwiners at the vertices, which ensure that the state transforms as
a singlet at the vertices. However, for this article, we will confine the discussion to the gauge
covariant coherent state, and in the conclusion, comment on the complications which can arise in a gauge invariant
state.

To end the discussion of the coherent state for gravity, we emphasise the following
two labels:\\
1)The classical phase space label $g_e$. These variables satisfy all the constraints by construction. Despite the discretisation involved through the definition of a graph, the variables respect the inherent continuity in the classical metric.\\
2)The semi-classicality parameter $t$.\\
The actual Coherent state is defined over the tensor product of Hilbert spaces, which for asymptotically flat manifolds can require a infinite tensor product of Hilbert spaces \cite{sto}.\\

\section{Semiclassical Black holes}
\subsection{The classical phase space}
\label{susec:clas}
The classical phase space for the spherically symmetric sector in gravity, will be constituted by Schwarzschild black holes. Clearly, one already knows the spherically symmetric classical solution of Einstein's equation in vacuum, and for this metric, all the constraints are satisfied. The intriguing part is to isolate the graph dependence to derive the discrete
classical phase space of Equation (\ref{eq:disc}). What exactly would be an appropriate graph? Obviously whatever the graph, the classical discretisation has to be spherically symmetric. One very convenient
set of graphs is to take edges along the coordinate lines $r,\th,\phi$ with appropriate discrete
labels attached to them. This was what was done in \cite{adg1}. The dual polyhedronal decomposition is then comprised of spherical surfaces which intersect the edges at their middle points. The holonomy and the momenta are calculated in \cite{adg1}, and their behaviour analysed. The radial holonomy and the corresponding radial momentum have the expressions:
\be
h_{e_r} = \cos \left(\tau'\left\{\frac1{r_2^{1/2}} - \frac1{r_1^{1/2}}\right\}\right) -
\imath\g^1 \sin\left(\tau'\left\{\frac{1}{r_2^{1/2}} - \frac1{r_1^{1/2}}\right\}\right)
\label{eq:rad}
\ee
Where $r_1$ is the begining of the edge, and $r_2$ is the end of the edge. ($\tau'= \sqrt{r_g}/2$ for $\beta=1$). The holonomy
is ofcourse independent of the angular coordinates of the begining and the end of the edge. Though the holonomy depends on the extrinsic curvature which diverges at the singularity, there is no
such divergence in the regulated `holonomy'. 

\includegraphics[scale=0.7]{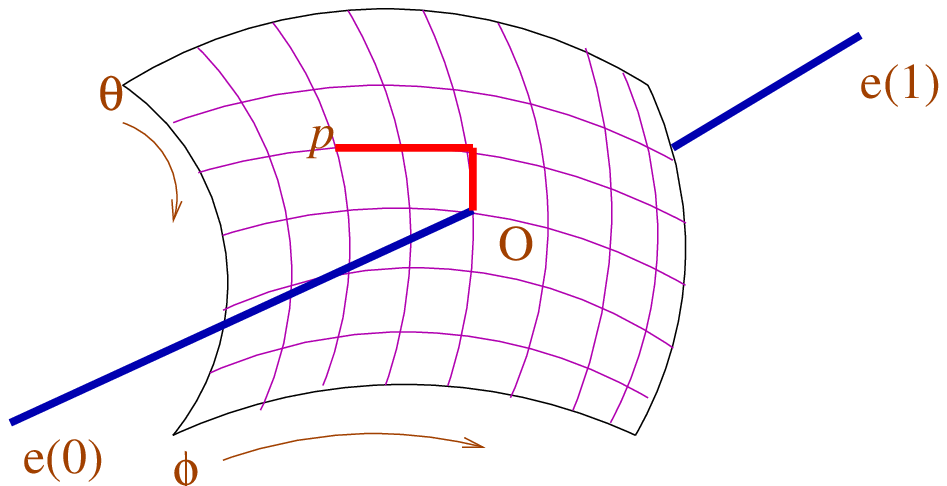}

The evaluation of the momenta is a little complicated, as it involves the evaluation of the integral on a two surface,
with the triad convoluted with the holonomies associated with paths from a generic point to the point at which the
edge intersects the two surface. As given in the above figure, the point of intersection of the edge with the surface
is denoted as O and has the coordinates (r,$\th_0, \phi_0$), \{in the final expressions $\phi_0$ does not contribute\}.
The width of the surface is from $\th_0-\th,\th_0 + \th$, and a linear dependence on the width in the $\phi$ direction,
which is suppressed here for brevity. The width along the radial edge is given by $\d$. For details see \cite{adg1},
\be
P^I_{e_r} = \frac{1}{a}Tr\left[\g^{I}\left(\cos\left(\frac{\tau'\delta}{2r}\right) - \imath \g^1\sin\left(\frac{\tau'\delta}{2r}\right)\right)X(r)\left(\cos
\left(\frac{\tau'\delta}{2r}\right) +  \imath\g^1\sin\left(\frac{\tau'\delta}{
2 r}\right)\right)\right] \label{momv2}
\ee
\be
X(r)= X_1(r)\g^1 + X_3(r) \frac{1}{\sqrt{\a^2 + 1}}\g^2  + X_3 \frac\a{\sqrt{\a^2 +1}}\g^3
\ee
with
\bea
X_1(r)&=& \frac{r_g^2}{\a^4} \sin\th_0 \left[ \frac{\sin(1-\alpha')\th'}{1-\alpha'} + \frac{\sin( 1 + \alpha')\th'}{ (1 + \alpha')}\right] \label{int}\\
X_3(r)&=& \frac{r_g^2}{\a^4}\cos\th_0 \left[ \frac{\sin(1-\a')\th'}{(1 - \a')} -
 \frac{\sin(1+ \a')\th'}{(1 + \a')}\right]  \label{int2}
\eea
Where $\a'= \sqrt{\frac{\a^2 +1}2}$, $\a=\sqrt{\frac{r_g}{r}}$.
Now, one uses the above in (\ref{momv2}), to get the following result for the momentum components
\bea
P^1_{e_r} = \frac{ X_1(r)}{a} &&  P^2_{e_r} =   \frac{X_3}{a\sqrt{\a^2 +1}} \left[\a
\sin\left(\g'\a^3\right) +\cos\left(\g'\a^3\right)\right]  \\ &&  P^3_{e_r} =  \frac{X
_3}{a\sqrt{\a^2+1}}\left[- \a\cos \left(\g'\a^3\right) + \sin\left(\g'\a^3\right
)\right] \nn
\label{momb}
\eea
($\g'=\d/2r_g$).
\bea
P_{e_r} &= & \frac{1}{a}\sqrt{P_1^2 + P_2^2 + P_3^2} = \frac{1}{a}\sqrt{X_1^2 + X_3^2} \label{momq}\\
&=&
\frac{r_g^2}{a~\a^4}\left[ \left(\frac{\sin[(1-\a')\th]}{(1-\a')}\right)^2 +
\left(\frac{\sin[(1+\a')\th]}{(1+\a')}\right)^2 \right.\nn\\ &-& \left.2\cos(2\th_0)
\frac{\sin[(1-\a')\th]}{1-\a'}\frac{\sin[(1+\a')\th]}{1+\a'}\right]^{1/2} \nn
\eea
Also, this complicated dependence on the coordinate point $\th_0$ at which the edge
intersects the dual surface becomes clear, when one takes the size $\th$ of the
surface to be very fine.
One notes that as the graphs get finer, the above approximates
$\th\rightarrow 0$ (restoring the width $\phi$):
\be
P_{e_r} = 2\frac{r^2}{a} \sin\th_0\th \phi.
\ee

Which is the area of the 2-surface which a edge intersects at it's middlepoint. Clearly, the
classical value of area is given as above. Question is, when one lifts the variable
$\sqrt{P^I_e P^I_e}$ to an operator, is it diagonalised in a area eigenstate? The answer is
yes, in a exact orthonormal eigenstate, this corresponds to the area operator and has the
eigenvalue $\sqrt{j (j+1)} t$. However, when we evaluate the expectation value in the coherent
state, classical area is corrected from this exact eigenvalue, and in the limit $t\rightarrow 0$
this has a different spectrum as we observe in Section (\ref{sub:equispace}). 
Before one
discusses the coherent state defined for these variables, some crucial points need to be noted:\\
1)The Holonomy and the Momenta remain finite numbers even in the vicinity of the singularity.
This can be attributed to a regularisation achieved due to the discretisation.\\
2)The variables are continuous across the horizon, and contain information about the apparent horizon.\\
3)For very fine graphs, one obtains as $P$ the classical area of the bit of the dual surface, induced
by the classical metric.\\
\subsection{The Coherent State}
The coherent state for the black hole phase space can be written explicitly as in \cite{tow}. The states are
peaked at the classical values, and are well behaved in the entire black hole slice. Due to the fact that the classical variables $h_e, P_e$ are themselves well behaved, there is no large fluctuations in the quantum corrections, at the horizon, or even at the vicinity of the singularity.
This is a indication of the `minimum uncertainty' principle obeyed by the states. The deviation around the mean value of both
the momentum, and the holonomy are measured by the variance, and they are such that
\be
<\psi^t|\Delta h_e ~\Delta P_e^I |\psi^t> = \frac{T^I}{2} h_e
\label{eq:un}
\ee
which being the classical holonomy, by Equation (\ref{eq:rad}) is always within the value $-1...+1$ in magnitude. Hence, the Coherent
states, are ideal to study the semi-classical regime, and due to the `regularisation' achieved in the
set of variables, here, the semiclassical approximation is valid in the entire black-hole slice.
The coherent state in the configuration space representation has the following expression
\be
\psi^t(gh^{-1}) = \sum_{j}(2 j +1)  e^{-t j(j+1)/2}\chi_j(gh^{-1})
\ee
where $j$ labels the eigenvalue of the SU(2) casimir, and $\chi_j$ corresponds to the character
of the corresponding irreducible representation. The corresponding momentum representation, is
determined by taking a `Fourier transform' in the character space of the irreducible states,
and one obtaines
\be
<\psi^t|jmn>= e^{-t j(j+1)/2}\pi_j(g)_{mn}
\ee
These are the position and momentum representations of the coherent state peaked at individual
edges, with the classical values encoded in the SL(2,C) valued variable $g_e$.
In this paper we make a very interesting observation for the state which is 'peaked' at the
classical value of $P=0$ or area =0, with arbitrary holonomy. For this state, the
expectation value of the area operator is however non-zero, and proportional to $t$,
as expected from the minimum uncertainty criteria demonstrated in equation (\ref{eq:un}).

\subsection{The Equispaced Area spectrum}
\label{sub:equispace}
There appears to be two different questions regarding the spectrum of
the area operator.\\
1)The eigenvalue of the area operator as obtained in an exact eigenstate.\\
2)The spectrum of the area of the black hole horizon, as measured in a appropriate
quantum state or a semiclassical state.\\

Regarding the first question, the usual regularisation of the area operator
\cite{rovsmo} gives the eigenvalue of the area operator
intersected by a edge to be : $8 \pi \sqrt{j (j+1)} l_p^2$ in a kinematical
eigenstate of the same operator \cite{rovsmo}.
In this context, in \cite{alek} it is claimed that the SU(2) casimir,
whose square root is proportional to the area operator undergoes
a {\it renormalisation} and gives the area {\it eigenvalue}  as $(j+1/2)l_p^2$, in it's
eigenstate \cite{alek}.
a
 Previous computations of entropy , 
counted the degrees of freedom of a boundary Chern Simons theory
at the horizon, given that the area of the horizon assumed certain values
as measured in exact eigenstates associated with edges crossing the horizon.
These edges as consistent with the previous calculations induced the horizon
with bits of area $8\pi \sqrt{j(j+1)} l_p^2$. 

The calculation which explains the semi-classical processes like black hole
entropy or Hawking radiation should arise from a appropriate semi-classical
limit of a quantum theory. To recover `semi-classical' entropy one needs to find
the microscopic degrees of freedom corresponding to a given {\it classical}
black hole space-time. The only well known states in a quantum theory which
give classical physics are the coherent states.
So, in a coherent state, the classical horizon area should be the `expectation value'
of an area operator. 
As discussed in Section(\ref{susec:clas}), the variable $P$ corresponds
to classical area. It is interesting to study the momentum representation of the coherent state
wavefunction, which is expanded in the area eigenstates to determine the classical area
 as a function of the area eigenvalue or the SU(2) Casimir eigenvalue $j$.
\begin{equation}
|\psi> = \sum_j d_j e^{-t j(j+1)/2}\pi_j(g)_{mn} |jmn>,
\end{equation}
where  $\pi_j$ is the jth irreducible representation of $g$ which is an SU(2) valued matrix encoding information about the
classical
variables. The state $|jmn>$ is an exact area operator eigenstate.
The coherent state is a superposition of such eigenstates with the coefficient
being the spin j irreducible representation of $g$. Now, the
probability distribution is,
\begin{equation}
\frac{ e^{-t j(j+1)}|\pi_j(g)_{mn}|^2}{||\psi||} \propto \exp( -\frac1t ((j+1/2)t -P)^2 ),\label{eq:prob1}
\end{equation}
where $P$ is the classical area. Hence, the classical area is
$P= (j_{cl} +1/2)t$, with maximum probability. It is also the expectation
value of the operator $\sqrt{\hat A + t^2/4}$ in an exact eigenstate, where
$\hat A$ is the usual area operator. Thus the classical area has a
'corrected
spectrum' determined by the {\it equispaced} discrete
numbers $(j_{cl} +1/2)t$ \cite{adg1,adg2}. Note that the word `spectrum' has to be used with caution as 
this is actually the expectation value of the area operator in a coherent state.
More precisely, given that the area operator is $\hat A$, the expectation value of the operator in the
coherent state is of the form:
\be
<\hat A>= <\psi|\hat A|\psi> = \frac1{||\psi||^2}\sum_j \sqrt{j(j+1)}t \ (2j+1) \bar\pi_j(g)_{mn} e^{-tj(j+1)}\pi_j(g)_{jmn}
\ee
Using $g_e=h_e e^{-iT^IP^I/2}$ and $||\psi||^2= \frac{2\sqrt{\pi}\sinh P}{t^{3/2} P} e^{-P^2/t} e^{t/4}$ and $\sum\pi_j(g^{\dag}g)=\chi_j(H^2)$, 
$H=e^{-i T^I P^I/2}$, $\chi_j(H^2)= \sinh(2j+1)P/\sinh P$.
the following can be derived 
\bea
<\hat A> & = & \frac{t^{3/2}e^{-P^2/t}e^{-t/4}\sinh P }{2 P \sqrt \pi}\sum_j \sqrt{\frac{(2j+1)^2t^2}{4} -\frac{t^2}{4}}\ (2j+1) e^{-t(2j)(2j+2)/4} \sum_{mn}\pi_j(g^{\dag})_{nm}\pi_j(g)_{mn} \nn \\
&=& \frac{\sqrt t \sinh P}{2P\sqrt{\pi}} \sum_j \sqrt{\frac14(2j+1)^2 t^2 -\frac{t^2}{4}} \ (2j+1) t \ \exp\left(-\frac{[(2j+1)^2t^2- 4 P^2]}{4t}\right)\frac{\sinh(2j+1)P}{\sinh P} \nn \\
&=& \frac{\sqrt t}{ 8 P \sqrt \pi} \sum_{n=-\infty}^{\infty}\sqrt{n^2 -t^2} \ n \ \exp\left(-\frac{(n-2P)^2}{4t}\right) , \ \ n=(2j+1)t \nn \\
&=& \frac{1}{4 P (2\sqrt {t\pi})} \int \sqrt{x^2 - t^2} \ x \ \exp\left(-\frac{(x- 2 P)^2}{4t}\right)  dx \nn \\
&=&\frac{1}{ 4 P}\int \sqrt{x^2} \ x \ \delta(x- 2 P)   dx  \ \   ({\rm Limit \ t\rightarrow 0}) \nn \\
&=& P  
\eea
Now, when the sum is converted to a integral in a variable x (step 3 of above), the corrections are proportional to $t$, and hence in first
order in $t$, this is a perfectly valid result \cite{abr} . However, for even a detectable finite $t$, this
result is true iff $2 P= (2 j_{cl} + 1/2)t$, or $P= (j_{cl} + 1/2)t$, where $j_{cl}$ are
discrete numbers.  
This observation is not contradictory to the previous work on area
spectrum of loop
quantum gravity, where the measurement is in an exact eigenstate. 
Note, in the $t\rightarrow 0$ limit, to assign a finite area, one has to take
 $j_{cl} >>1$, and hence the equispaced spectrum assigns a continuous value 
 as the non-equispaced spectrum : $j_{cl} t$ to the area. 

When one is trying to
count the number of ways to build a macroscopic area, using the coherent states, the classical
areas induced by the edges, are counted by the discrete numbers $j_{cl}$. There is a degeneracy
associated with every $j_{cl}$, corresponding to the numbers $m,n =-j_{cl}..j_{cl}$ (corresponding to the expectation
values of the operators $P^3_R, P^3_L$), and this is $2j_{cl}+1$. This degeneracy gives a degeneracy associated with  
 a given horizon area. Why this should be the entropy
of a black hole is the subject of discussion in the next few sections. In the next subsection
however, we discuss the resolution of the singularity at the center of the black hole.
We will now concentrate on the specific coherent state, whose classical label has
contributions only from the unitary component of $g_e$, namely the holonomy, and the $P^I=0$, or the classical area induced by the particular edge is 0.
(or $g_e=h_{cl}$). In this particular choice of graph, this will happen for areas induced by edges close to $r=0$. 
 The coherent state is:
\be
|\psi^t_0 (h_{cl})> = \sum_{jmn}d_{j} e^{-t j(j+1)/2}\pi_j(h_{cl})_{mn}|jmn>
\ee
 
In the above $g_e= h_{cl}$, as the $P^I=0$.
The expectation value of the area operator in the coherent state is
\bea
<\psi^t_0(h_{cl})|\hat P |\psi^t_0(h_{cl})>&& \nn \\
& = & <\psi^t_0(h_{cl})| \sum_{jmn} d_j\sqrt{j(j+1)}t e^{-t j(j+1)/2} \frac{\pi_j(h_{cl})_{mn}}{||\psi||}|jmn>\\
& = & \sum_{j mn} \sqrt{j(j+1)}t e^{-t j(j+1)}\frac{\pi_j(h_{cl}^{\dag} h_{cl})_{mn}}{||\psi||^2} \label{eq:area1}\\
&=& \frac{1}{||\psi||^2}\sum_{j} \sqrt{j(j+1)} td_j e^{-j(j+1)t} \chi_j(1)\\
& = & \frac12 t + O(t^2)
\label{eq:area0}
\eea
In equation (\ref{eq:area1}), one uses $<jmn|klq>= \frac{1}{d_j}\  \d_{jk}\d_{ml}\d_{nq}$, and $\bar\pi_{j}(h_{cl})_{mn}= \pi_j(h_{cl}^{\dag})_{nm}$
 
 Any measurement on the coherent state gives the classical expectation value only
when $t\rightarrow 0$, even a tiny amount of $t$ would ensure that there is minimum
area which one can measure in the coherent states. Thus one obtaines in some sense a minimum
radius which can be measured in the spherically symmetric coordinates, 
\be
<\hat P> = \frac{2}{a} r_{min}^2\sin\th_0 \th\phi
\ee
for each edge. Which gives the value of $\frac{r_{min}^2}{a} = \frac n{ 8 \pi} t$, where $n$ is the number of edges inducing the total area of the sphere. Thus even for a infinitesimal $t$, and fine graph, there is a minimum radius and this uncertainty leads to a
resolution of the singularity as we shall subsequently observe.
\subsection{The Curvature operator and the Singularity}
The Curvature scalar $R_{\m\n\l\s}$ will consist of contributions from the intrinsic metric,
as well as contributions from the extrinsic curvature. These are of the form:
\be
R_{\m \n \l}^{\s} = q^a_{\m} q^b_{\n} q^c_{\l} q_d^{\s} R^{d}_{abc} - K_{\m \n}K_{\l}^{\s} + K_{\n\l}K_{\m}^{\s}
\ee
Clearly, when the intrinsic metric is taken to be flat, one has the curvature scalar as
\be
R_{\m\n\l}^{\s} R^{\m \n \l}_{\s} = 2\left[ K^4 - K_{bc}K^{ac}K^{d}_a K^b_d\right]
\label{ext}
\ee
(In principle there can be order $t$ corrections to the intrinsic curvature, but to first order,
the contribution is essentially zero. Hence we ignore the intrinsic curvature terms here.)
This quantity diverges at $r=0$ classically for the Schwarzschild metric. One can write the entire above
expression in terms of the regularised holonomy and momenta. To see the form, we write the
expression for the extrinsic curvature in the form:
\be
K_{ab} = \frac{1}{\b}\left[\G^I_a - A^I_a\right] E_{b I} \sqrt{\rm det{E}}
\ee
Further, the spin connection $\G^I_a$ is written in terms of the momenta:
\bea
\G^I_a &= &\frac12 \e^{IJK} E^b_K\left[ E^J_{a,b} - E^J_{b,a} + E^c_J E^L_{a} E^L_{c,b}\right] \nn\\
&+ &\frac14 \e^{IJK} E^b_K\left[2 E^J_a \frac{ (\rm det E)_{,b}}{\rm det E} - E^J_b\frac{(\rm det E)_{,a}}{\rm det E}\right]
\eea
The entire curvature is then finally written in terms of the two operators
$h_e(A), P_e(A)$ and their expectation values in the coherent states.
The $R^2$ operator is
\bea
R^2 &= &2 \left[\left\{\left(\G^I_a - A^I_a\right)\left(\G^I_b - A^I_b\right)\frac{E^a_I E^b_I}{\rm \det E}\right\}^2 \right. \nn \\ & - &\left. \left(\G^I_a - A^I_a\right)\left(\G_{Ib} - A_{Ib}\right) \left(\G^k_c - A^k_
c\right) E^d_k \left(\G^l_d -A_d^l\right)E_l^b \left(E^a_l E^c_l\right)({\rm \det E}^2)\right]
\label{eq:curv}
\eea
Before one actually lifts the above expression to a operator equation $h_e,P_e$, one must also express
inverse powers of triads which appear in the
Equation(\ref{eq:curv}), in terms of Poisson Brackets.
Thus a measurement of $R^2$ in the coherent state will be obtained after one has replaced the
$A_a^I$ and the $E_a^I$ in terms of the holonomy and the corresponding momentum, and the regularised
expression for the inverse triads $E^a_I$. {\it In the vicinity of the singularity, the
terms containing $\G^I_a$ donot contribute to the singularity of the curvature
but all go to zero at $r=0$, and hence can be ignored in the calculation for the upper bound
of the curvature operator}. It is the terms containing $A_a^I$ 
which are potentially divergent. Thus in the curvature operator, we retain the terms independent of the
spin connection. Note, as discussed later for the apparent horizon equation,
it is possible to write the extrinsic curvature operator solely in terms of the gauge connection operator,
and the triad operator, by using the Immirzi parameter (\ref{eq:ext12}). Hence all the derivations for 
the extrinsic curvature, here will be true, when one takes into considerations the appropriate $\b$ of 
the theory. 
However, for the next few discussions, we
`ignore' the spin connection in the computation of the curvature operator in the vicinity 
of the singularity. Since the quantum fluctuations are always small, one can never induce
large values for the spin connection operator, when their classical value is 0. Thus for a measurement
of the curvature operator in the coherent state, in the vicinity of the singularity, it is always
justified to ignore the spin connection operator.

Before evaluating the expectation value of the curvature operator,
we make the following observations on the operator ordering ambiguity which occurs
for the operators which are functions of both $h_e$ and $P_e^I$. These will clarify some of our
assumptions and calculations. 

In field theories, operator ordering ambiguities often lead to infinities. However, for these
coherent states, the ambiguities are proportional to $t$ and should go to zero in the
classical limit. This observation might not be true when the classical geometry itself has a singularity. 
To investigate the situation where the classical geometry itself is singular, we take a arbitrary function 
of $ f = h P$ with, the normal ordering defined as
\bea
<\psi|:f:|\psi> & = & <\psi| (P^I_e) h_e|\psi> + <\psi|[(P^I_e), h_e]|\psi> \nn \\
&=& (P^I_e h_e)_{cl} + \frac{t}{2}~ T^I (h_e)_{cl} + O(t^2)
\eea
Clearly, the operator ordering ambiguity is proportional to $h_{e cl}$. 
Let us examine the radial case,
Clearly, since by Equation (\ref{eq:rad}), we find that taking one of the end points of the
edges to $r_1(r_2)\rightarrow 0$,

\bea
h_{e_r} = \cos\left(\frac{\tau'}{\sqrt{r_1}}\right)\cos\left(\frac{\tau'}{\sqrt{r_2}}\right) &+ &
\sin\left(\frac{\tau'}{\sqrt{r_1}}\right)\sin\left(\frac{\tau'}{\sqrt{r_2}}\right)  \nn \\
+ \imath\g^1\left[\cos\left(\frac{\tau'}{\sqrt{r_1}}\right)\sin\left(\frac{\tau'}{\sqrt{r_2}}\right)\right.
& - &\left.\sin\left(\frac{\tau'}{\sqrt{r_1}}\right)\cos\left(\frac{\tau'}{\sqrt{r_2}}\right)\right]
\eea
 the limits $\cos(1/\sqrt{r_1(r_2)}), \sin(1/\sqrt{r_1(r_2)})$ oscillate
within the finite limits $-1..1$. Hence in terms of the `regularised variables' taking the
limit to the singularity does not affect the coherent state, and as such, indeed the
coherent state can be defined in terms of the regularised classical variables, even in the
vicinity of the singularity. Thus now any arbitrary polynomial in terms of the holonomy and momenta, 
can be obtained as a expectation value and classical limit determined even at the singularity. 
Thus we proceed to write the curvature operator in terms of the holonomy and momentum, and 
find the expectation value. 
 
It is a interesting calculation to realise how the classical singularity can be recovered from a
`finite regulated' holonomy, 
For example, with Equation (\ref{eq:rad}), the radial gauge field e.g. will be:
\be
A^I_r = -\frac{1}{2(r_1-r_2)} {\rm Tr}[T^I (h_{e_r} -1)] 
\ee
Where $r_1$ and $r_2$ are the begining and end points of the edge respectively.
Now the most general form of the holonomy can be taken as $h_{e_r}= e^{T^I\s^I}= \cos\sigma + \frac{T^I\s^I}{\s}\sin\s$,
which implies
\bea
A^I_r &= & -\frac{1}{2(r_1 - r_2)}{\rm Tr}[\cos \sigma T^I - \frac{\s^I}{\s}\sin\s + \frac{\e^{IJK}\s^JT^K}{\s}\sin\s]\nn\\
&=& \frac{1}{r_1 - r_2}\frac{\s^I}{\s}\sin\s \nn \\
\eea
Here the interesting observation is that now in the particular value of holonomy calculated above:
$\sigma^I\propto (r_1^{-1/2} - r_2^{-1/2})\delta^{I1}$, which by itself is divergent in the limit
$r_1,r_2\rightarrow 0$.  $\sigma^1/\sigma=1$, and the $\sin\s$ factor infinitely
oscillates from -1..1 as $\s$ is diverges. Thus, even in the value of $A_r^I$, the divergence
does not appear to show up, in the regularised variables or the extrinsic curvature
of the manifold if there exists a minimum edge length. However, it is the question of taking limits, 
and if somehow, one shrinks the edgelength faster than the limit $r\rightarrow 0$, then the
$\sin\s$ factor can be approximated as $\s$, and the following divergence appears:
\bea
&=& \frac{\sqrt r_g}{r_1-r_2}\left(r_1^{-1/2} - r_2^{-1/2}\right) \nn \\
&=& \frac{\sqrt r_g}{\sqrt{r_1r_2}\left(\sqrt{r_1} + \sqrt{r_2}\right)}
\eea
In the limit $r_1\rightarrow r_2$ and $r_2\rightarrow0$, this is indeed a singular value
of the extrinsic/gauge connection recovered. Thus it is very clear that in terms
of the regularised variables, even the existence of a minimum edge length gives a upper
bound on the curvature proportional to $\frac{1}{\d^4}$, where $\delta$ is the minimum
edge length. As described above for a non-zero value of $t$, there is a minimum area
one can measure in the coherent states, which would also give a bound on the length of
the edges as measured. However, we will attempt to measure the
curvature directly using instead the regularised extrinsic curvature operator, and demonstrate a possible
resolution of the singularity in terms of the coherent states. First let us approximate 
$K_{ab}$, (ignoring $\G^I_a$) using the usual regularisation as follows:
\be
K_{ab} = - A^I_a e^I_b
\ee
Using the previous regularisations of the inverse triad, one can obtain a suitable expression as follows: 
Writing $e^I_b= C Tr[T^I h^{-1}_{e_b}\{h_{e_b},V\}]$, note that $C$ is a graph dependent constant,
which is necessary, as the volume operator, defined in terms of the $P_I's$ is graph dependent. 
In fact as in \cite{hanno}, the constants are fixed here by calculating the Poisson bracket of the
holonomy operator with the volume operator. 
Since $V= \sqrt{\frac{1}{3!} \e^{IJK}e_{abc} P^a_I P^b_J P^K_c}$, the Poisson bracket of the holonomy
with the Volume operator is:
\bea
\{h_{e},V\}&=& \{h_e,\sqrt{\frac1{3!} \e^{IJK}\e_{abc} P^a_I P^b_J P^c_K}\} \nn \\
&=& \frac1{2 V} \{h_e, \frac1{3!} \e^{IJK}\e_{abc} P^a_I P^b_J P^c_K\} \nn \\
&=& \frac\k{8a V} \e^{IJK}\e_{abc}  h_{e_a} T^I P^b_J P^c_K \nn \\
&=& \frac{\k V}{4a}  h_{e_a} T^I(P^a_I)^{-1} \nn \\
&=& \frac {\k V}{4 a} h_{e_a} T^I P_a^I \nn \\
&=& \frac{\k v}{4 s_{e_a} a} h_{e_a} T^I E_a^I \sqrt{q}\nn \\
&=& \frac{\k v}{4 s_{e_a} a} h_{e_a} T^I e_a^I \nn 
\eea
here, $v= \th \phi\d/a^{3/2}$ and $s_{e_r}=\th\phi/a,s_{e_\th}= \d\phi/a, s_{e_\phi}= \d \th/a$.
($\d,\th,\phi$ denote the edge lengths along the coordinate directions). Thus, now multiplying by $ T^I h_{e_a}^{-1}$ and taking Trace, gives
the constant to be
\be
 C_{e_a}= \frac{ 2 a s_{e_a} }{ v \k}
\label{const}
\ee

In the quantum commutator, $C\propto 1/l_p^2$.
\be
K_{ab} = C_{e_b} Tr[T^I\frac{h_{e_a}}{2(e(1)-e(0))}]Tr[T^I h_{e_b}^{-1}[h_{e_b},V]]
\label{eq:quant}
\ee

\be
<\psi^t|K_{ab}|\psi^t> = C_{e_b} <\psi|Tr[T^I\frac{h_{e_a}}{2(e(1)-e(0))}] Tr[T^I (V- h^{-1}_{e_b} V h_{e_b})]]|\psi>
\ee
However, to avoid the presence of the double trace in the operator and one instead uses the following 
regularisation of the extrinsic curvature:
\bea
K_{ab}& = -& A_a^I e_b^I  \\
& = & \frac{C_{e_b}}{2(e_a(0) - e_a(1))}{\rm Tr} \left[ h_{e_a} h_{e_b}^{-1} \{h_{e_b}, V\} \right]
\label{eq:regt}
\eea
($e_a(0), e_{a}(1)$ denote the begining and end point of an edge).
With the classical value recovered in the limit the edge length goes to zero.
The $h_{e_a}$ denotes the holonomy along the edge $e_a$, and $V$ is the corresponding volume operator. The Poisson
brackets give the inverse triads and the limit the edge lengths go to zero, one is left with the classical
expression for the extrinsic curvature. However, when one lifts the above to an operator equation, the operator
ordering is taken to be:
\be
K_{ab} = \frac{C_{e_b}}{2(e_a(0)- e_a(1))}{\rm Tr}\left[ h_{e_b}^{-1}[h_{e_b}, V] h_{e_a}\right]
\label{reg}
\ee
As this ensures that the diagonal components are recovered appropriately. 
For the radial component of the extrinsic curvature, one obtains:
\be
K_{rr} =\frac{C_{e_r}}{2(e_r(1)-e_r(0))} {\rm Tr}\left[ h_{e_r}^{-1}[h_{e_r}, V]h_{e_r} \right]
\ee

Thus a typical term in the evaluation of the curvature will include the Volume operator.
To find the spectrum of the volume operator in these coherent states, we have to
derive the operator in more details.

\subsection{Volume Operator}
As in contrast to previous derivations of the volume operator, here, there exists a classical
metric, to fix the constants and the exact expression for the operator in terms of the gauge
invariant vector fields. 
We will proceed in the following manner:
Firstly, we fix the classical volume in terms of the graph degrees of freedom, and then
lift the obtained expression to a operator in the quantum theory. The volume operator
will be obviously adapted to the specific graph chosen. Here we will also focus on
a set of vertices in a local region R. The dual polyhedronal decomposition of the
manifold will be important in the determination of the individual volume cells.
Here, the graph has been taken to be 6-valent with three ingoing and three outgoing
edges at a given vertex. Since the classical intrinsic metric has been taken in the
spherical coordinates, the edges are aligned along the coordinate directions. As described
before the triads are smeared over the dual 2-surfaces which the edges intersect at their
midpoints. The dual surfaces constitute a volume cell, with a vertex at the center of the
cell. This geometric construction very conveniently follows from the definition \cite{thiem}.
The figure shows clearly the construction of the volume cell, and each vertex is surrounded
by the dual surfaces forming the walls. The volume associated with each vertex is therefore
\be
V= \sqrt{\frac1{3!} \e^{a b c}\e_{IJK} P^I_{e_a} P^J_{e_b} P^K_{e_c}}
\ee
Where $e_a, e_b, e_c$ are a triplet of edges intersecting at the vertex. For a generic vertex
located at the point (as measured by the classical metric ) $(r,\th_0,\phi_0)$, the classical
volume as evaluated from above can be evaluated. Typically,
For the edges ingoing at the vertex, contribute with (\ref{momb})[ for the momentum of angular edges \cite{adg1}],
\be
P^1_{e_r} = (r-\d)^2 \sin(\th_0-\th/2) \frac{\th \phi}{a} \ \ \ \ \ P^2_{e_\th}= (r-\d/2) \sin(\th_0-\th)\frac{\d\phi}{a} \ \ \ P^3_{e_\phi} = (r- \d/2)\frac{\d\th}{a}
\ee
 The contribution from the outgoing edges are similarly of the form:
\be
P^1_{e_r}= (r + \d)^2 \sin(\th_0 + \th/2)\frac{\th \phi}{a} \ \ \ \ P^2_{e_\th}= (r+ \d/2) \sin(\th_0 + \th)\frac{\d\phi}{2 a}  \ \ \ \ \ P^3_{e_\phi} =(r+ \d/2)\frac{\d\th}{2 a} 
\ee
The determinant to first order in the edges of the cube ($\d, \th,\phi$) is obtained as:
\be
V= r^2\sin\th_0 \frac{(\d \th\phi)}{a^{3/2}}
\ee
Which is the required volume of the cell.

\includegraphics[scale=0.8]{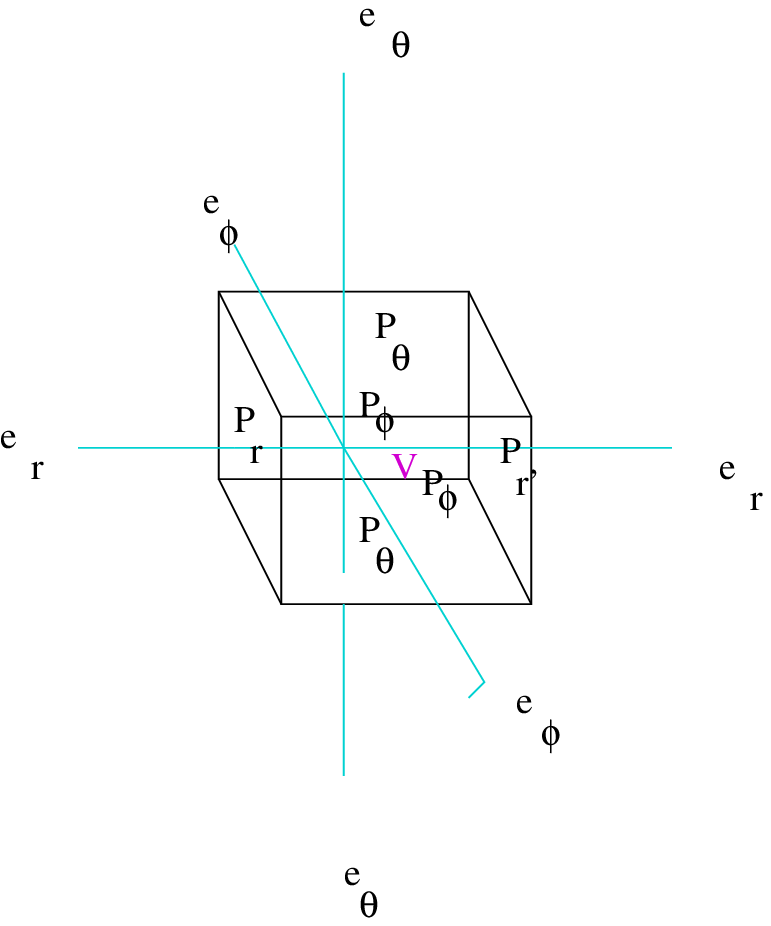}

Thus the entire volume of the manifold can be obtained as the sum of the volume of the cells. 
The spectrum of the volume is then obtained in the coherent state by using the standard techniques.
The operator $P^I_e$ is replaced by the Right invariant vector fields $X^a_I$ and then the expectation
value of the operator is obtained in the state. This operator is of course quite similar
to previous derivations, however, here the basic cell is prompted by the classical metric, and the
six-valent graph adapted to that.
Now, to evaluate explicit matrix elements, in the coherent states, one obtaines the following:
\be
<\hat V^2> = \frac{1}{3!}<\psi|\e^{abc}\e_{IJK} X^a_I X^b_J X^c_K|\psi>.
\ee
This operator, again due to a trick due to Thiemann, can be written as:
\be
\e_{IJK} X^a_I X^b_j X^c_k = [ (X^a+ X^c)^2, (X^b + X^c)^2]
\ee
which however is not a convenient set of operators at this juncture, as we are interested
mainly in the classical limit, where a naive evaluation of the operator expectation values
gives 0, (recall that due to the nature of the classical metric, $<X^a. X^b >$ =0).
Instead, we use the sperical symmetry of the classical metric to equate the following:
\be
{\rm Limit}_{t\rightarrow 0}< \hat V> = {\rm Limit}_{t\rightarrow 0}<\hat P_r> \frac{\delta}{\sqrt{a}}
\ee
as $P_r = r^2\sin\th_0 \th \phi$ has the same magnitude as $\sqrt q$. This would be true only at the first order
in the metric, and the result will be considerably different in higher order corrections
in the semiclassicality parameter $t$. However, since we are interested in a plausible
resolution of the singularity, we approximate the volume operator at this level of the
discussion by the above.

\subsection{Resolution of Singularity}

To find the expectation value of the curvature operator in the coherent state, we evaluate the
explicit expressions of the complete operator.
The operator is taken as a density 1 operator as this gives some nice properties.
Thus, the operator is of the form:
\be
\sqrt{g} R^2
\ee
by construction. This as by equation (\ref{ext}) now has the form (including only the potentially diverging terms):
\bea
&&2 \sqrt{q} \left[ K^4 - K_{ab}K_{cd}K^{ac} K^{bd}\right] \\
&& 2 \sqrt{q} \left[ (K_{ab} K_{cd} q^{ac} q^{bd}) ^2 - K_{ab}K_{cd} K_{ef} K_{gh} q^{ae}q^{fc} q^{bg} q^{gh}\right] 
\label{extr}
\eea

The regularisation of the extrinsic curvature is then used as from equation (\ref{eq:regt}), however, the inverse metric
$q^{ac}$, when regularised, introduces inverse powers of ${\rm det} q$, which are then absorbed into the Poisson brackets
in the numerator, and also with a point splitting method introduced by Thiemann.
This essentially has the following method, which uses the fact that:
\bea
1 & = & \frac{\rm det e_a^I}{ \sqrt{q}}  \nn \\
&=& \frac{1}{3! \sqrt q}\e^{IJK}\e^{abc} e^a_I e^b_J e^c_K \nn \\
&=& \frac{C_a C_b C_c}{3! \sqrt q} \e^{IJK} \e^{abc} Tr\left[T^I h_{e_a}^{-1} \{h_{e_a}, V\}\right] Tr\left[ T^J h_{e_b}^{-1}\{h_{e_b}, V\}\right]Tr\left[T^K h_{e_c}^{-1}\{h_{e_c}, V\}\right]
\label{point}\eea
The above Poisson brackets then add to the brackets with the volume, and the inverse powers in the denominator can then
be absorbed in the numerators. We concentrate in the first term of Equation (\ref{extr}), and an explicit expression with the extrinsic curvatures
is the following:
\bea
\sqrt{q} K^4 
= (q^{1/4} K ^2)(q^{1/4} K^2)
= ( \frac{\rm det e_a^I}{q^{1/4}} K^2)^2
\eea
Now, we find using the point splitting of equation (\ref{point}), and the regularisation of the Extrinsic curvature
of Equation(\ref{reg}):
\bea
&& \frac{\rm det e^I_a}{q^{1/4}} K^2 \nn  \\
&=& \frac{C_{e_m} C_{e_n} C_{e_p} C_{e_a} C_{e_c}}{3! q^{1/4}}\e^{IJK} \e^{mnp} Tr\left[T^I h_{e_m}^{-1}\{h_{e_m}, V\}\right] Tr\left[T^J h_{e_n}^{-1}\{h_{e_n}, V\}\right] Tr\left[T^K h_{e_p}^{-1}, \{h_{e_p}, V\}\right] \nn \\ 
&\times & \frac{1}{e_b e_d s_{e_a} s_{e_b} s_{e_c} s_{e_d} }Tr[h_{e_a}^{-1}\{h_{e_a}, V\} h_{e_b}] Tr\left[h_{e_c}^{-1}\{h_{e_c}, V\} h_{e_d}\right]\frac{ P^{a}_M P^{b}_M P^{c}_N P^{d}_N}{q^2} 
\eea
where by $e_b, e_d$ we denote the length of the edges along those directions.
 The numerator has five Poisson brackets, whereas the denominator has the power of volume
$q^{9/4} = v^{9/2}V^{9/2}$, which when absorbed in the five brackets, gives a contribution of
$ \frac{1}{V^{9/10}}\{h_e, V\}=\frac{1}{10}\{h_{e}, V^{1/10}\}$ for each. 
Thus now the first term of the operator in Equation (\ref{extr}) has the form:
{\small \bea
&& \frac{(10)^{10}(C_{e_m} C_{e_n} C_{e_p} C_{e_a} C_{e_c})^2}{v^{9}s_{e_a}^2 s_{e_b}^2 s_{e_c}^2 s_{e_d}^2 e_a^2 e_b^2}\left[\frac{1}{3!} \e^{IJK}\e^{mnp} Tr\left[ T^I h_{e_m}^{-1}\{h_{e_m}, V^{1/10}\}\right]Tr\left[ T^J h_{e_n}^{-1}\{h_{e_n}, V^{1/10}\}\right] \right. \nn \\ &\times& Tr\left[T^K h_{e_p}^{-1}\{h_{e_p},V^{1/10}\}\right] 
\left.Tr\left[ h_{e_a}^{-1}\{h_{e_a}, V^{1/10}\} h_{e_b}\right] Tr\left[ h_{e_c}^{-1}\{h_{e_c}, V^{1/10}\}h_{e_d}\right]P_M^aP_M^b P_N^c P_N^d\right]^2
\label{exp}
\eea}
Now, the subsequent task is to find the expectation value of the operator in the coherent states,
by first lifting the Poisson brackets to commutators. Thus now rather lengthy expressions occur,
though the calculation is quite straightforward. One of the main observation in the evaluation
of the operators which are product of quite a few terms is the fact that one can insert
the complete set of coherent states which are resolutions of unity. Though the states peaked at different
$g$ are not orthogonal to each other, the overlap function vanishes as $t\rightarrow 0$, and hence
the contribution to first order in $t$ will be from the terms which are expectation values
at the same $g$. In other words to quote a theorem in \cite{tow}, $<\psi(g)|\psi(g')>$ is exponentially
vanishing as $t\rightarrow 0$ ($d \mu_L$ is the Liouville measure). 
\bea
{\rm Limit}_{t\rightarrow 0}<\psi| XY|\psi> &&  \nn \\
& = & {\rm Limit}_{t\rightarrow 0}\int d\mu_L(g')<\psi|X |\psi'(g')><\psi'(g')|Y|\psi> \nn  \nn \\
& = & {\rm Limit}_{t\rightarrow 0}[<\psi|X|\psi><\psi|Y|\psi> \\ &+&\left(\int<\psi|X|\psi'><\psi'|Y|\psi>-<\psi|X|\psi><\psi|Y|\psi>\right)] \nn \\
& =& <\psi|X|\psi><\psi|Y|\psi>  + O(0)
\eea
Thus, one now, can take the terms in (\ref{exp}), and then one inserts the complete set of coherent states
so as to isolate the individual trace terms. The expression in (\ref{exp}), is a sum of products of
trace terms. Thus the expectation value of the product of the operators can be broken into product of 
expectation values as follows:
\bea
&& <\psi|\sqrt{q}K^4|\psi>\nn \\
&=&\left[\frac{C}{Q}<\psi|\frac{1}{3!}\e^{IJK}\e^{mnp}Tr\left[T^I h_{e_m}^{-1}\{h_{e_m},V^{1/10}\}\right] Tr\left[T^J h_{e_n}^{-1}\{h_{e_n},V^{1/10}\right]\right. \nn \\
&\times & \left.Tr\left[T^K h_{e_p}^{-1}\{h_{e_p},V^{1/10}\}\right]Tr\left[h_{e_a}^{-1}\{h_{e_a},V^{1/10}\},h_{e_b}\right]Tr\left[h_{e_c}^{-1}\{h_{e_c},V^{1/10}\}h_{e_d}\right]P^a_M P^b_M P^c_N P^d_N |\psi>\right]^2 \nn \\
&=&\left[\frac{C}{Q}\frac{1}{3!}\e^{IJK}\e^{mnp}<\psi|Tr\left[T^I h_{e_m}^{-1}\{h_{e_m},V^{1/10}\}\right]|\psi><\psi|Tr\left[T^J h_{e_n}^{-1}\{h_{e_n},V^{1/10}\}\right]|\psi>\right. \nn \\
&\times&<\psi| Tr\left[T^K h_{e_p}^{-1}\{h_{e_p},V^{1/10}\}\right]|\psi><\psi|Tr\left[h_{e_a}^{-1}\{h_{e_a},V^{1/10}\}h_{e_b}\right]|\psi> \nn \\
&\times & \left. <\psi| Tr\left[h^{-1}_{e_c}\{h_{e_c},V^{1/10}\} h_{e_d}\right]|\psi><\psi|P^a_M P^b_M P^c_N P^d_N |\psi>\right]^2 
\label{equation}
\eea
($C= 10^{10}(C_{e_m} C_{e_n} C_{e_p} C_{e_a} C_{e_c})^2$, and $Q= v^9 s_{e_a}^2 s_{e_b}^2 s_{e_c}^2 s_{e_d}^2 e_a^2 e_b^2$
Thus, once the expectation values of the individual trace terms are taken, a typical term in 
the expansion is of the form (\ref{equation}):
\bea
&& <\psi| {\rm Tr}\left[T^I(V^{1/10} - h_{em}^{-1}V^{1/10}h_{em})\right]|\psi> \nn \\
& = &-\sum_A<\psi|{\rm Tr}\left[(T^I h_{em}^{-1} V^{1/10} h_{em})\right]_{AA} |\psi>
\label{exp2}
\eea

The first term in the above vanishes due to the presence of the trace of $T^I$, which is 0.
From the results of \cite{tow1}, one can directly replace the expression by their classical values,
however, since the coherent states are not eigenstates of the holonomy operator, we take a more careful
approach in our analyses of the curvature operator. One proceeds by taking $\hat h_{ AB} = e^{-3t/8} [e^{i \hat P^I T_I/2 }\hat g]_{AB}$ and
with $\hat h_{AB}^{-1} = e^{-3t/8} [\hat g^{\dag} e^{i \hat P^I T^I/2}]_{AB}$. The coherent states are eigenstates of the
operators $\hat g$ and $\hat g^{\dag}$ and hence one can extract their eigenvalues from the expectation value. The equation (\ref{exp2}) gives:
\be
\sum_{ABB'CC'} T^I_{AB} e^{-3t/4} g^{\dag}_{BB'} g_{C'A} <\psi| e^{i \hat P_I T_I/2}_{B'C} V^{1/10} e^{i \hat P^J T^J/2}_{CC'}|\psi>
\ee

Here $A,B =0,1$ denote the SU(2) index.
Using the Baker-Campbell-Hausdorff formula, the quantity in the brackets assumes the form:
\be
e^{t/4} <\psi|\left[\cosh\frac p2 -  \frac t2 \frac{\sinh p/2}{p} + iT^I P_I \frac{\sinh p/2}{p}\right]V^{1/10}\left[\cosh(p/2) - \frac t2 \frac{\sinh p/2}{p} + i T^I P_I \frac{\sinh p/2}{p}\right]|\psi> 
\ee
where the operator $p=\sqrt{P^2 + t/4}$. It is interesting to note that this operator never has the
zero eigenvalue, even in the coherent state peaked at the classical area $P=0$. 
The sum now reduces to 
\bea
&& \sum_{ABC} T^I_{AB}g^{\dag}_{BC}g_{CA}e^{-t/2}<\psi|\left[\cosh(p/2) - \frac t2\frac{\sinh p/2}{P}\right]V^{1/10}\left[\cosh(p/2) -\frac t2 \frac{\sinh p/2}{p}\right]|\psi> \nn \\ 
& + & e^{-t/2}\sum_{ABB'C} T^I_{AB} g^{\dag}_{BB'}T^J_{B'C}g_{CA}e^{-t/2}<\psi|\left[P^J\frac{\sinh p/2}{p} V^{1/10}\left(\cosh\frac p2 - \frac t2\frac{\sinh p}{p}\right)\right]|\psi> \nn \\
& + & e^{-t/2}\sum_{AB'C'} T^I_{AB} g_{BC}T^J_{CC'}g_{C'A}<\psi| \left(\cosh\frac p2 - \frac t2\frac{\sinh p}{p}\right)(P^J)\frac{\sinh p/2}{p}|\psi> \nn \\
& - & \sum_{ABB'C'A}e^{-t/2}T^I_{AB}g^{\dag}_{BB'}T^J_{B'C}T^K_{CC'}g_{C'A}<\psi|  P^J \frac{\sinh p/2}{p} V^{1/10} P^K \frac{\sinh p/2}{p} |\psi>
\label{equations2}
\eea 
The above, clearly in the limit $t\rightarrow 0$ takes the classical values, but which are
bounded as $r\rightarrow r_{min}$. Now, the operator whose expectation value is
to be evaluated is not a potentially divergent term, in the vicinity of the singularity,
which is essentially $P\rightarrow0$. Any divergence, shall be in the terms including
$g,g^{\dag}$, due to their dependence on the classical holonomy.
To isolate the potential divergent terms in the above, we simplify each of the traces in the 
terms in the above.

The first trace term to simplify in (\ref{equations2}) is:
\bea
&&{\rm Tr} (T^Ig^{\dag} g)  \nn \\
&\approx &{\rm Tr} (T^I h^{-1} e^{-i P^I T^I/2} e^{-i P^I T^I/2} h) e^{3t/4} \nn \\
&=& {\rm Tr} (h T^I h^{-1} e^{-i P^I T^I}) e^{3t/4} \nn \\
&=&{\rm Tr}(T^I e^{-i P^I T^I})e^{3t/4} + e^{3t/4}{\rm Tr}([h,T^I] h^{-1} e^{-i T^I P^I}) \nn \\
&=& i 2 P^I \frac{\sinh P}{P} e^{3t/4} - e^{3t/4}\frac{2 \s^J\sin\s}{\s}\e^{IJK} {\rm Tr}(T^K h^{-1} e^{-i T^I P^I}) \nn \\
&\approx& 2 i P^I \frac{\sinh P}{P} e^{3t/4} - e^{3t/4}\frac{2 \s^J\sin \s}{\s} \e^{IJK}\left[2 i\frac{P^K}{P}\sinh P \cos\s\right. \nn \\&&\left. + 2\s^K\frac{\sin\s}{\s}\cosh P
- 2i \e^{LMK}\frac{\s^{L}P^{M}}{\s p}\sin\s \sinh P\right]
\label{eq:tr1}
\eea
For the next few equations, we concentrate on the Trace terms one by one, which containes the $h$ dependence
and hence a potential divergence term. 
\bea
{\rm Tr}(T^I g^{\dag} T^J g) && \nn \\
&=& {\rm Tr}(T^I h^{-1} e^{-i T^I P^I/2} T^J e^{-i T^IP^I/2} h) e^{3t/4}\nn \\
&=& {\rm Tr}(h T^I h^{-1} e^{-i T^I P^I/2} T^J e^{-i T^I P^I/2}) e^{3t/4}\nn \\
&=& {\rm Tr}(T^I e^{-i T.P/2}T^Je^{-i T.P/2})e^{3t/4} - \e^{IJK}\frac{\s^J\sin\s}{\s}\left[T^K h^{-1} e^{-i T.P/2} T^J e^{-i T.P/2}\right] e^{3t/4}\nn \\
&=& {\rm Tr}(T^I e^{-i T.P/2} T^J e^{-i T.P/2})e^{3t/4} -\e^{IJK}\frac{\s^J\sin\s}{\s}\left(\cos\s{\rm Tr}(e^{-i T.P/2}T^Je^{-i T.P/2}T^K)\right. \nn \\&&\left. -\s^L\frac{\sin\s}{\s}{\rm Tr}(T^Le^{-i T.P/2}T^J e^{-i T.P/2}T^K)\right) e^{3t/4}
\label{eq:tr2}
\eea
The next trace term is quite similar. The other term is then equal to
\be
{\rm Tr}(T^I g^{\dag} T^J T^K g)= -{\rm Tr}(T^I g^{\dag} \d^{IJ} g) + \e^{IJM}{\rm Tr}(T^I g^{\dag} T^m g)
\label{eq:erd}
\ee
The first term in (\ref{eq:erd}) is of the form in equation (\ref{eq:tr1}), and the second term is as in
equation (\ref{eq:tr2}).
Thus the dependence on the holonomy would be precisely of the
form as stated earlier. 
Clearly, all the terms of the above operator are bounded as $P\rightarrow 0$, which is the
location of the singularity. 
The other type of the terms as obtained from equation (\ref{equation}) are:
\be
<\psi|Tr\left[h^{-1}_{e_a}\left[h_{e_a}, V^{1/10}\right] h_{e_b}\right]|\psi><\psi|{\rm Tr}\left[h^{-1}_{e_c}\left[h_{e_c},V^{1/10}\right] h_{e_d}\right]|\psi><\psi|P_M^aP^b_M P^c_NP^d_N|\psi>
\ee
The terms for arbitrary $a,b$ lead to extremely complicated terms, which we refrain from writing here,
as the specifics are not important, and the observation of the non-divergence is continued for those
terms also. See Appendix B for the case of the non-diagonal metric. What we discuss here, is where the classical metric is assumed to be the diagonal
spherically symmetric one, and hence $P_M^a P_M^b= (P^a)^2\d^{ab}$ and hence, one eventually gets
a simplified set of terms, one a factor of which would be:
\be
<\psi|Tr\left[V^{1/10}h_{e_a} - h^{-1}_{e_a}V^{1/10}h_{e_a}h_{e_a}\right]|\psi>
\label{eq:exp}
\ee
The first term when written in terms of $g_{AB}$, would simplify to
\be
\sum_{A} e^{t/4}<\psi| V^{1/10}(\left[\cosh(p/2) - \frac t2 \frac{\sinh p/2}{p} + i T^JP^J \frac{\sinh (p/2)}{p}\right]_{AB} g_{BA}|\psi>
\ee
The first two terms simply include the trace terms of $g$ and the last term is of the form:
\be
{\rm Tr} (T^I g)= \left(\cos\s \cosh P + \frac{\s^IP^I}{\s}\sin \s \sinh P\right)e^{3t/8}
\ee

The next term would be considerably more complicated, but the procedure would be the same,
and the terms obtained would be of the following form:
\be
-Tr<\psi|h^{-1}_{e_a}V^{1/10}h_{e_a}^2|\psi>
\ee
which is 
\be
\sum_{A,B,C,D,E,F} e^{t/4} g^{\dag}_{e_a \ AB}<\psi|e^{iT^IP_I^a}_{BC}V^{1/10} [e^{iT^J P_J^a}]_{CD} g_{e_a DE}[e^{i T^J P_J^a}]_{EF}g_{e_a FA}|\psi>
\ee

From here the trace terms would be of the form:
\bea
{\rm Tr}(g^{\dag} g g)e^{-9t/8}& = & {\rm Tr}(h^{-1} e^{-iTP/2} e^{-iTP/2} h e^{-iTP/2} h)\nn \\
&=& {\rm Tr}(e^{-iTP}he^{-i TP/2})= {\rm Tr}(he^{-i 3/2 TP})\nn \\
&=& \cos\s Tr(e^{-i 3/2 TP}) + \frac{\s^I\sin\s}{\s}{\rm Tr}(T^I e^{-i3/2 T.P}) 
\eea

Then:
\bea
{\rm Tr}(g^{\dag} T^I g g) & = & {\rm Tr}(h e^{-i TP} T^I e^{-i TP/2}) \nn \\
&=& \cos\s {\rm Tr}(e^{-3 T.P/2}T^I) + \frac{\s^J\sin\s}{\s}{\rm Tr}(T^J e^{- i TP}T^I e^{-i TP/2}) 
\eea

Next:
\bea
{\rm Tr}(g^{\dag} gT^I g) & = & {\rm Tr} (e^{- 3i TP/2} hT^I) \nn \\
&=& {\rm Tr} (h T^I e^{- i 3 TP/2})= \cos\s {\rm Tr}( T^I e^{- i 3TP/2}) + \frac{\s^J\sin\s}{\s}{\rm Tr}(T^J T^I e^{-i 3 TP/2}) 
\eea
Finally:
\bea
{\rm Tr} (g^{\dag} T^I g T^J g) &=& {\rm Tr}(h T^J e^{- iTP} T^I e^{- iT P/2}) \nn \\
&=& \cos\s Tr (T^J e^{- i TP} T^I e^{-i TP/2}) + \frac{\s^K\sin\s}{\s}{\rm Tr}(T^K T^J e^{-i TP} T^I e^{-itP/2})
\eea

The second term in Equation (\ref{ext}) is of the form:
\be
 - \sqrt{q}\left[K_{ab}K_{cd} K_{ef} K_{gh} q^{ae}q^{fc} q^{bg} q^{gh}\right] 
\ee
This has to be written in terms of the holonomy and the momentum, using the expressions in (\ref{eq:quant}),
which is:
\bea
&&- \sqrt{q}\frac{C_{e_a} C_{e_c} C_{e_e} C_{e_g} }{e_b e_d e_f e_h s_{e_a} s_{e_b} s_{e_c} s_{e_d} s_{e_e} s_{e_f} s_{e_g} s_{e_h}}\left[{\rm Tr}\{h_{e_a}^{-1}[h_{e_a},V]h_{e_b}\}{\rm Tr}\{h_{e_c}^{-1}[h_{e_c},V]h_{e_d}\}{\rm Tr}\{h_{e_e}^{-1}[h_{e_e},V]h_{e_f}\}\right.\nn \\ &\times& \left. {\rm Tr}\{h_{e_g}^{-1}[h_{e_g},V]h_{e_g}\} 
 \frac{P^a_IP^b_I P^c_JP^d_J P^e_KP^f_K P^g_L P^h_L}{q^{4}}\right] \nn \\
&=& - \frac{C_{e_a} C_{e_c} C_{e_e} C_{e_g} }{e_b e_d e_f e_h s_{e_a} s_{e_b} s_{e_c} s_{e_d} s_{e_e} s_{e_f} s_{e_g} s_{e_h}}\frac{\rm (det e_I^a)^2}{q^{9/2}}\left[{\rm Tr}\{h_{e_a}^{-1},[h_{e_a},V] h_{e_b}\}{\rm Tr}\{h_{e_c}^{-1}[h_{e_c},V]h_{e_d}\}{\rm Tr}\{h_{e_e}^{-1}[h_{e_e},V]h_{e_f}\} \right. \nn \\ &\times& {\rm Tr}\{h_{e_g}^{-1}[h_{e_g},V]h_{e_h}\}
 \left. P^a_I P^b_I P^c_J P^d_J P^e_K P^f_K P^g_L P^h_L\right]\\
&=& - \frac{C_a C_c C_e C_g C_m C_{m'} C_n C_{n'} C_p C_{p'} v^{9}} {e_b e_d e_f e_h s_a s_b s_c s_d s_e s_f s_g s_h}\frac{1}{(3!)^2}\e^{IJK}e^{mnp} {\rm Tr}[ T^I h_{e_m}^{-1}[h_{e_m},V^{1/10}]] {\rm Tr}[T^J h^{-1}_{e_n}[h_{e_n},V^{1/10}]] \nn \\ &\times& {\rm Tr}[T^K h^{-1}_{e_p}[h_{e_p},V^{1/10}]] \e^{I'J'K'}\e^{m'n'p'}{\rm Tr}[T^{I'}h^{-1}_{e_{m'}}[h_{e_{m'}},V^{1/10}]]{\rm Tr}[T^{J'} h_{e_{n'}}^{-1}[h_{e_{n'}},V^{1/10}]] \nn \\ &\times& {\rm Tr}[T^{K'} h_{e_{p'}}^{-1}[h_{e_{p'}},V^{1/10}]]  
\left[{\rm Tr}\{h_{e_a}^{-1},[h_{e_a},V^{1/10}] h_{e_b}\}{\rm Tr}\{h_{e_c}^{-1}[h_{e_c},V^{1/10}]h_{e_d}\}{\rm Tr}\{h_{e_e}^{-1}[h_{e_e},V^{1/10}]h_{e_f}\} \right. \nn \\ &\times& {\rm Tr}\{h_{e_g}^{-1}[h_{e_g},V^{1/10}]h_{e_h}\}
 \left. P^a_I P^b_I P^c_J P^d_J P^e_K P^f_K P^g_L P^h_L\right]
\label{exp21}
\eea
The evaluation of the expectation value of the above operator will follow in the precise way as
discussed above. The expressions would correspond to the evaluation of (\ref{eq:exp}). 
Thus now all the terms which appear in the expectation value of the curvature operator have been
expressed as functions of regularised variables which are completely bounded.
The curvature operator is a function of a particular vertex and the edges meeting at that point.
If the vertex is limitingly taken to $r\rightarrow 0$ or $P\rightarrow 0$ in the regularised variables, then the curvature
is always finite if the edge length is kept as non-zero. All the expectation
value of the operators are finite. From a analysis of the constants
in front of the expression for Equation (\ref{exp}) and in Equation (\ref{exp21}), the
terms go as $\frac{1}{t^{10}a^{1/2}e^3}$, where $e$ is a typical edge length.
Even if all the edge lengths are infinitesimal, the divergence reappears
iff $t\rightarrow 0$. The reason is that, the expectation value of the
volume operator, $V^{1/10}$ contributes from each term, to give finally
the expression $\frac{V}{ t^{10} e^3\sqrt a}$. Thus taking the edgelengths to
0 also takes the Volume to zero in the numerator. 
The divergence is prevented absolutely by the observation that there is a
minimum area as long as there is a non-zero $t$. 
This resolution of the curvature singularity is a calculation of the expectation value of the 
actual operator in coherent states and different than calculations of the inverse scale factor
in cosmological theories \cite{cosmo}.

To summarise:\\
1)The Kretschmann scalar $\sqrt g R^2$ is taken and it's expectation value evaluated in the given 
coherent states.\\
2)The operator is written in terms of the extrinsic curvature (\ref{extr}), as by construction, the coherent states
are peaked on a spatial slice whose intrinsic curvature is 0.\\
3)The terms in the Extrinsic curvature are regularised in terms of the holonomy and the dual momenta as in
equation(\ref{eq:quant}).\\
4)The calculation of the expectation value of the operators gives finite terms and a proportionality
to powers of $1/t$. This shows that the singularity reappears as $t\rightarrow 0$.
 
\section{The Apparent Horizon}
We now proceed to the question of origin of black hole entropy. 
 Clearly we are dealing with a single spatial slice here, and have
not tried to follow the evolution of the black hole space-time dynamically. The coherent
state has also been constructed to obtain the geometry of a spatial slice. So,
one cannot talk about global quantities like the Event Horizon, and even `Isolated
Horizons' and try to obtain boundary conditions on the coherent state wavefunction by studying their 
pull backs on the given spatial slice. Hence, the relevant
quantity to study is the Apparent Horizon, where the knowledge about the intrinsic metric
and the extrinsic curvature of the slice are enough to determine the existence of a trapped
surface of the equation. Thus, the geometry of the slice is built using a coherent state,
such that it includes the apparent horizon, and then one can proceed to integrate out the
wavefunction which is inside the apparent horizon, and obtain a suitable `entropy of the
apparent horizon'. In the following discussion, we shall talk about apparent horizons which
have $S^2$ topology. Given a coherent state, we try to re-build the classical
space-time by evaluating the expectation values of momenta and the holonomy in
these states. We recover the information about the extrinsic curvature and the
intrinsic metric of the slices. Question is how does one know that there is a
apparent horizon in the slice? There is a very general equation, which when satisfied
shows the existence of a trapped surface. This is of the following form:
\be
\nabla_a S^a + K_{ab} S^a S^b - K=0
\ee
Where $S^a$ is a space-like vector, normal to the 2-surface.
The above can be re-written in terms of the variable $K^I_a$ and the triads
as:
\be
\nabla_a S^a + K_{a}^Ie_{b}^I S^a S^b - K_a^I e^a_I=0
\ee
For arbitrary value of the Immirzi parameter, this equation will show a dependence
on the parameter. Here, as $ \ ^{\b}K_{a}^I= \b K_a^I$ and $ \ ^{\b} e_a^I= (1/\sqrt\beta) e_a^I$,
 $\ ^{\b} e^a_I= \sqrt{\beta} e^a_I$, the above equation in terms of the generelised variable
assumes the form (the Christoffel connections which appear in the covariant derivative are
independent of $\beta$):
\be
\nabla_a S^a + \sqrt\beta \ ^{\b}K_a^I \ ^{\b}e_b^I - \frac1{\b^{3/2}} \ ^{\beta}K_a^I \ ^{\beta} e^a_I=0
\ee

Now, we stay with the assumption that we are trying to re-build the Schwarzschild
black hole space-time from the coherent state, and hence, there is a set of
spherically symmetric coordinates. The apparent horizon, is a sphere or
$S^a=(1,0,0)$, and one gets the following relation for the intrinsic metric measured, and
the extrinsic metric of the same slice. (We resort to the classical variables or $\beta$=1).
\be
 - \G^{\th}_{\th r} -\G^{\phi}_{\phi r} + K_{rr} - K_{rr} q^{rr} - K_{\phi\phi} q^{\phi\phi} - K_{\th \th} q^{\th \th}=0
\label{eq:app}
\ee
Now, $\G^{\th,\phi}_{\th,\phi r}, K_{\th,\th}, K_{\phi,\phi}$ in the classical values cancel each other at
$r=r_g$ leaving the following equation in the radial sector, which is trivially satisfied
everywhere:
\be
K_{rr} (1-q^{rr})=0
\ee
Thus clearly imposing the apparent horizon equation on the radial edges does not
introduce any new constraints on the radial coherent state wavefunction. One can then
think of the horizon as being formed by purely radial edges crossing the horizon, with
the radial wavefunction corresponding to `free wavefunctions' as would be there for
a spherically symmetric space time.

\includegraphics[scale=0.8]{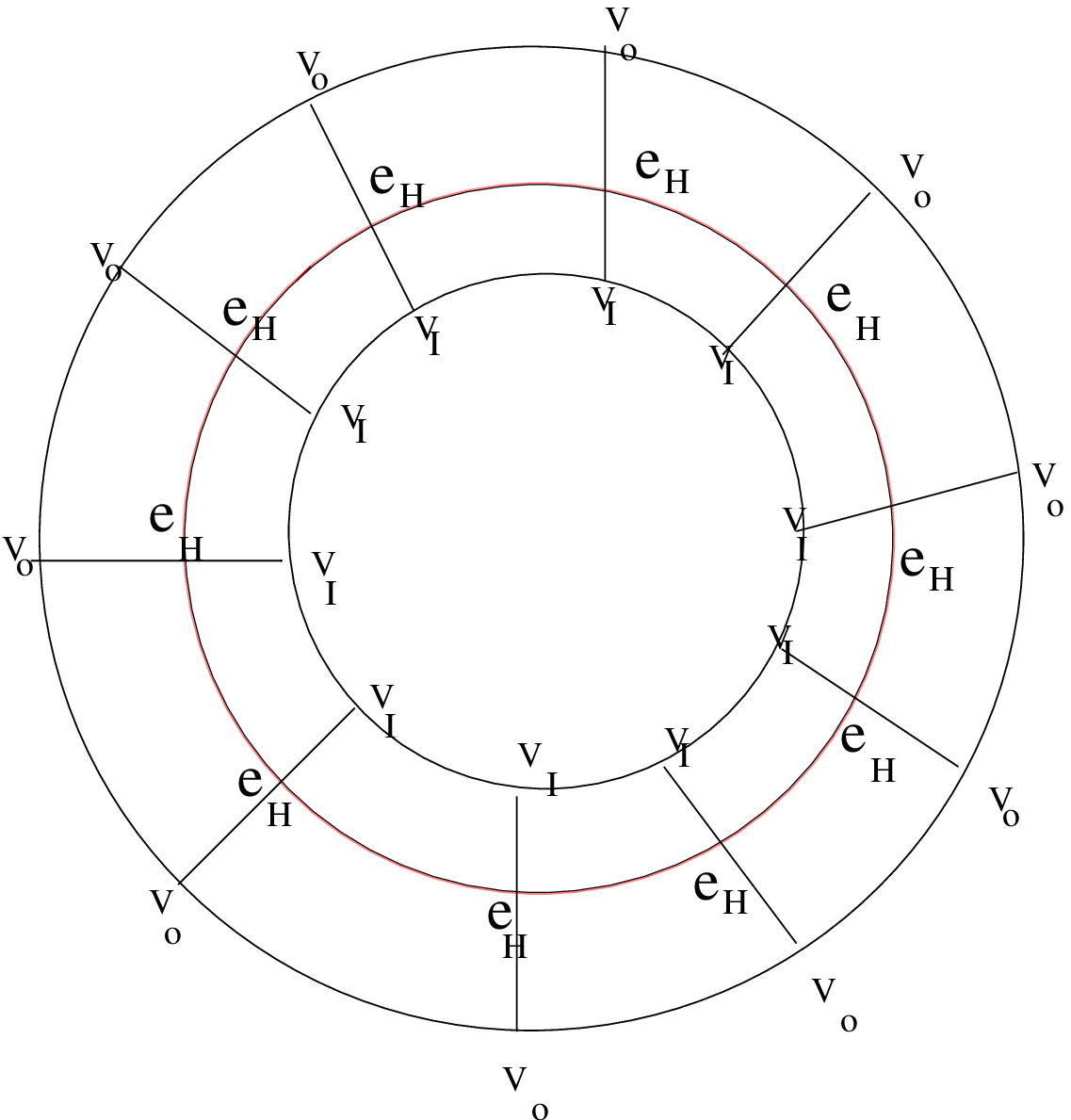}

While trying to lift the apparent horizon equation to an operator equation, we 
only retain the terms which constrain the angular sector of the theory. 
A crucial assumption which is included in the calculation is the fact that 
classical spatial slice has a flat spherically symmetric metric, and the horizon
is a spherical surface in that spatial slice. 

The apparent horizon equation now only includes the derivatives of the angular metric forming a
difference equation for the discretised variables along the angular edges. The equation is:
\be
-\G^\th_{\th r} - K_{\th}^I e^{\th}_I - \G^{\phi}_{\phi r} - K^I_{\phi} e^{\phi}_I =0
\label{eq:app1}
\ee
The $\G^{\th}_{\th r} = \frac12 q^{\th \th}(q_{\th \th})_{, r}= e^{\th}_I e^{\th}_I (e_{\th J  \ ,\  r} e_{\th J})$
The derivative is replaced by a difference in the discrete version of the
same equation. The horizon, is thus now a set of radial edges, with a set of vertices $\{v_1\}$ outside the
horizon, and a set of vertices $\{v_2\}$ inside the horizon. The derivatives are thus differences in the value
of the metric at vertices $\{v_1\}$, with their values which are at vertices $\{v_2\}$. Thus for a representative
set of vertices, 
the difference equation is obtained by subtracting the
value of $e_{\th J}$ at $v_1$ which is outside the horizon with the value of the inverse
dreibein inside the horizon at vertex $v_2$.
 The inverse dreibein is also expressed as a Poisson bracket first
and then lifted to an operator form. Thus now

\bea
\G^{\th}_{\th r} & = & e^{\th}_I e^{\th}_I \frac1{r(v_1) - r(v_2)}(e_{\th}^J(v_1) - e_{\th}^J(v_2))e_{\th}^J(v_1) \nn \\
&=& \frac{P^{\th}_I P^{\th}_I}{ \left(r(v_1) - r (v_2)\right) V^2}\left[ Tr\left[ T^J h_{\th}^{-1}\left[h_{\th},V\right]\right]_{v_1} - Tr\left[ T^J h_{\th}^{-1}\left[h_{\th}, V\right]_{v_2}\right]\right] \nn \\
& \times & Tr\left[ T^J h_{\th}^{-1}\left[h_{\th}, V\right]\right]_{v_1}  
\eea

Further, one uses the observation that $^{\beta} A_{a}^I= \Gamma_a^I - \b K^I_a$, where $\b$ is the 
Immirzi parameter, to write $- \beta\partial_{\b} \ ^{\beta}A^a_I=  ^{\beta}K^I_a$. This enables one to rewrite the
apparent horizon equation interms of the holonomy operator, incorporating information of the
extrinsic curvature.
Thus 
\be
^{\beta}K_{\th}^I \ ^{\beta}e^{\th}_I = -\beta\partial_{\b} Tr[T^I \ ^{\beta}h_{\th}] \frac{^{\beta}P^{\th}_I}{V}
\label{eq:ext12}
\ee
The equation is then of the form :
\bea
&& \frac{P^{\th}_I P^{\th}_I}{V}\left[Tr\left[ T^J h_{\th}^{-1}[h_{\th}, V]\right]_{v_1} - Tr\left[T^J h_{\th}^{-1} [h_{\th},V]\right]_{v_2}\right]Tr[ T^J h_{\th}^{-1}[h_{\th},V]]_{v_1} \nn \\
& & \ \ \ \ \ -\frac{1}{\sqrt \beta}\frac{\partial}{\partial {\beta}} Tr[T^I \  ^{\beta} h_{\th}] \ ^{\beta}P^{\th}_I + (\th \rightarrow \phi)= 0 \nn \\
&{\rm or}& 4 P^{\th}_I P^{\th}_I \left[Tr\left[ T^J h_{\th}^{-1}[h_{\th}, V^{1/2}]\right]_{v_1} - Tr\left[ T^J h_{\th}^{-1}[h_{\th}, V^{1/2}\right]_{v_2}\right] Tr[T^J h_{\th}[h_{\th}, V^{1/2}]]_{v_1} \nn \\
&& \ \ \ \ \ \ -\frac{1}{\sqrt \beta}\frac{\partial}{\partial \b} Tr[T^I \ ^{\b} h_{\th}] \ ^{\b} P^{\th}_I + (\th\rightarrow \phi) =0
\label{diff}
\eea

Since the first set of terms is independent of $\beta$ no attempt has been made to write the
$\beta$ index there.

Fortunately at the classical level, the difference equation has a linear term in $r$ to derive.
In other words, the derivative is of the form
$-\frac1{\delta} (r- (r+\delta))= r$, and hence yields a exact answer for the 
derivative, introducing no further fuzziness in the equation for the location of the apparent horizon, for fine graphs. 
An evaluation of the equation in the coherent states leads to the following equation:
\bea
&& \sum_{ABCDE}\left(T^J_{AB} g^{\dag \th}_{BC} g^{\th}_{EA}<\psi|\left(e^{i TP^{\th}/2}\right)_{CD} V^{1/2} \left(e^{i T P^{\th}/2}\right)_{DE}|\psi>\right) \nn \\ 
&- & \frac{1}{4\beta}\frac{\partial}{\partial\beta}<\psi|T^I \  ^{\beta} h_{\th}|\psi>\left. \frac{<\psi|P_I^{\th}|\psi>}{<\psi|P^{\th}_I P^{\th}_I|\psi><\psi|Tr(T^Jh^{-1}_{\th} V^{1/2} h_{\th}|\psi>}\right|_{v_1} \nn \\
&=& \left. \sum_{ABCDE} T^J_{AB} g^{\dag}_{BC}g_{EA}<\psi|\left( e^{i T P_{\th}/2}\right)_{CD} V^{1/2} \left(e^{i T P_{\th}}\right)_{DE}|\psi> \right|_{v_2}
\eea
Now, in the above equation, all the quantities on the LHS contain quantities which are functions of variables
defined inside the horizon, whereas, rest of the quantities are of variables defined outside the horizon. 
Now one can proceed and evaluate the traces like in the curvature operator case, and obtain complicated
expressions on both sides of the equation. The solution for the equation, exists and in the $v_1\rightarrow v_2$
limit gives $r=r_g$. Without
going into the explicit solution, one can infer some qualitative features which are sufficient for the
discussion of the entropy.\\
1) The quantisation introduces a `fuzzyness' with the location of the
horizon, which is proportional to the semiclassicality parameter.\\ 
2) the classical values of $g_{\th},g_{\phi}$
for the edges meeting at the vertex $v_1$ are correlated with the same for edges meeting
at the vertex $v_2$. In other words,
\be
g_{e(v_2)}\equiv g_{e(v_2)}(g_{e(v_1)})
\ee
3)The apparent horizon equation is completely independent of the holonomy along the radial edge
which crosses the horizon.\\
4)The Volume operator at the vertex $v_2$ contains information about the horizon Area $P_H$, 
but we consider that as an independent variable, which can be changed, and the rest of the
variables will change accordingly. This will imply a sum over all possible graphs interpretation
for the entropy, which we also try to discuss at the end of this article. For a fixed $P_H$,
the above apparent horizon equation, gives the graph degrees of freedom inside the horizon
as a function of those outside the horizon. These will reflect in a correlation between the
wavefunctions inside and outside the horizon.
To emphasise again, the radial edges which cross the horizon remain
completely free, with no restriction on them from the apparent horizon equation.

Clearly this results in a set of wavefunctions for the coherent state peaked
at the edges which are correlated with each other. Now, one has to ensure that
this results in a entropy proportional to the area of the horizon.
We show in the final density matrix calculation, it is the radial
edges which constitute the `entropy' of the system by counting the number of
ways to build the apparent horizon 2-surface area. 

In more general terms, very crucial
assumptions which went into the determination of Equation (\ref{eq:app1})
to have a separation of the angular and radial components was the assumption
that there are no cross terms in the extrinsic curvature, or
$K_{r\th(\phi)}=0$ even at the operator level. This can be ensured in terms of the gauge connection
by the condition
\be
K_{ab} = (\G^I_{a} - A^I_{a})e_{Ib} =0,  {\rm for}    a\neq b
\ee
and
\be
q^{ab} = E^a_I E^b_I=0 {\rm for} a\neq b
\ee
(Which is the diagonalisation condition of the metric).

However, these assumptions are not drastic, as any deviations from the above due
to quantum fluctuations will be proportional to $t$, and hence ignored in the
correlators being considered here. 
Thus, the apparent horizon is located by measurements in the angular coherent states,
and the radial coherent states associated with the radial edges, induce the apparent
horizon with area. Now, we shall address the question why the degeneracy associated
with the induction of the Area of the horizon can be called the entropy of the
black hole. Hence, in the next section, we first review, how entropy arises from
a density matrix, and then derive the density matrix associated with the
coherent state wavefunction.

\section{The density matrix calculation}
The density matrix by definition for a given configuration in a state
$|\psi>$ is of the following form:
\be
\rho= |\psi><\psi|
\ee
The expectation value of any operator can be evaluated in this Matrix
as :
\be
<A> = Tr(A\rho) = Tr(A|\psi><\psi|) = <\psi|A|\psi>
\ee
The usefulness of the density matrix is particularly evident, when the system
is in the product of two Hilbert spaces, and one defines a 'reduced'
density matrix by tracing over one of them. As a illustration, let the
system be in the state:
\be
|\psi> = \sum_{im} d_{im}|i>|m>
\ee
where the orthonormal states $|i>|m>$ have a product structure, with
$|i>$ and $|m>$ belonging to two different Hilbert spaces. One can define a
reduced density matrix, by tracing over the $|m>$ states.
\bea
\rho_{\rm {red}} &= & Tr_{m} \left[\sum_{im} d_{im}|i>|m> d^*_{jn} <j|<n|\right] \nn \\
&=& \sum_{im}\sum_{jn} d_{im}d^*_{jn} |i><j| \delta_{mn} \nn \\
&=& \sum_{ijm} d_{im}d^*_{jm} |i><j|
\label{eq:trace}
\eea
Interestingly, the diagonal elements of the density matrix are the probability
of finding the system in a state $|i>$, given any state in the Hilbert space
which has been traced over.
or,
\be
\rho_{\rm {red} ii} = \sum_{m} d_{im} d^*_{im}
\ee
and hence the condition which naturally arises is:
\be
Tr(\rho_{\rm{red}})=1
\ee
If there are no correlation between the Hilbert spaces, in particular
if the coefficients factorise as $ d_{im} = d_i d_m$, then the resultant
reduced system is still a pure system. This can be checked by the following
condition:
\be
\rho^2 = \rho \ \ \mbox{the State is Pure}, \ \ \ \ \rho^2 < \rho \ \ \ \mbox{the state is mixed}
\ee
 As reported in the previous section, the Coherent state for a black hole has
precisely the same structure of product over Hilbert spaces. Now, is there
a observer for whom only a part of the system is measurable? At the classical
level, (one believes that the measuring instruments are classical), for an outside
observer, the inside of a black hole does not exist. He/She does not receive
any signal from inside the horizon. For such an observer, there has to be a
reduced density matrix. However, as the coherent state is written in the covariant
form, there appear no information of the continuity conditions which must be ,
imposed at the vertices. Thus, the classical labels of different edges are not really independent, but must satisfy a continuity condition at a shared vertex. This introduces correlations even at the classical level. These continuity conditions must be imposed from Einstein's equation. Here, we discuss only the classical correlation introduced due to the apparent horizon equation. 
Thus even at the classical level, and one cannot integrate
over the edges inside the horizon independently of the edges outside. There is also
a method to quantify the relation.
\subsection{Correlations}
Classically, we have defined the graph degrees of freedom for the black hole, as the
holonomy and momentum along one-dimensional edges and corresponding dual surfaces.
Though, we have talked about recovering the `classical action' principle for these
variables on the individual edges, we have yet to examine the continuity conditions
at the vertices, since the classical metric itself is continuos throughout
the manifold. These conditions will ensure the existence of correlations for the
states defined on the entire manifold. At the level of the gauge invariant 'coherent state' wavefunction,
the Gauss constraints are satisfied with the introduction of the Clebsch Gordon coefficients
at the vertices. There is a discussion on the Gauge invariant states in \cite{tow}, and
the peak behaviour of these states. However, we are mainly interested in quantifying
the correlation at the vertices, and in the semiclassical limit. This limit is difficult
to evaluate from the exact form of the coefficients. We shall discuss this eventually,
but for the time being, we discuss the `macroscopic correlations' which 
are manifestations of the continuity of the classical metric, and at the horizon,
the apparent horizon equation.

For the classical observer, who makes measurements in the
Schwarzschild coordinates $t,r,\th,\phi$,  while confined to the slice $\tau=c$ slice,
the spatial slice ceases to exist after the horizon.
Though the transformed discrete variables will be the same set of variables of the
observer, as $\frac{dt}{d\tau}=1$, but as he/she approaches the apparent horizon,
for her/him, the coordinate time measured in $t$ will be running to infinity.
(This is the same graph, same slicing, but measured in the coordinates of the
outside observer, and not the $\tau,R,\th,\phi$ frame.)
For the  outside observer, the coherent state
wavefunction inside the horizon is irrelevant information. He/she
is forced to deal with a reduced density matrix instead. The expectation
values of all operators for the outside observer pertain to the density matrix now.
What does tracing over the `Inside edges' mean? The gauge-invariant
wavefunctions are correlated at the vertices by the Clebsch-Gordon coefficients.
And it is a difficult task to evaluate the exact nature of this correlation,
and how the coefficients influence the peak of the entire coherent state.
However, the fact that the elements $g$ themselves
are classical solutions restrict the wavefunctions. As mentioned in the
previous section, the apparent horizon equation clearly correlates
the classical elements across the horizon.
We now try to quantify the correlation across the horizon. Let us take
the set of radial edges which cross the horizon. These end at a set of
vertices, and these vertices are the starting point of a set of edges,
now inside the horizon. If the radial edges which cross the horizon induce
the set of classical elements $g=e^{i P^i T_i/2} h_e$, then the areas induced by the edges
inside the horizon also get determined from the apparent horizon equation. 

The apparently uncorrelated wavefunctions are now naturally correlated. Let us 
take the specific coherent states which are peaked over the edges sharing a vertex, 
one inside the horizon, and one outside the horizon. For simplicity, we will be suppressing
the azimuthal direction. The
edges are $e_{r_o}$, $e_{\th_o}e_{\th_o'}, e_{H}, e_{r_i}, e_{\th_i}, e_{\th_i'}$. Where the first
three share a vertex $v_1$ with the radial edge $e_H$ crossing the horizon, and the last three
sharing vertex $v_2$ which is inside the horizon and end point of $e_H$. To avoid overcounting of the edges sharing the vertex $v_1$ and $v_2$, one takes the set of angular edges which are ingoing at a given vertex. The coherent state is thus 
a product of the coherent states for each of the above edges.
\bea
|\Psi(v_1,v_2)> &= &\frac{1}{||\Psi||^2}  \sum_{j's} \pi_{j_o}(g_{r_0})\pi_{k_o}(g_{\th_0})\pi_{j_H}(g_H)\pi_{j_i}(g_{r_i})\pi_{k_i}(g_{\th_i}) \nn \\
&\times& e^{-t/2\sum_{j's}j's (j's +1)} |j_o>|k_o>|j_H>|j_i>|k_i>
\eea
The $j's $ notation denotes all the spin eigenvalues $j_o,....k_i$.
 
Now from the apparent horizon equation, 
classical values of $g_{\th_i}=g_{\th_i}(g_{\th_o})$, and hence
the the wavefunctions cannot be integrated as independent, for the different edges.
The complete wavefunction should be of the form:
\be
|\Psi(v_1,v_2)>= \sum_{j_{\{O\}}j_H j_{\{I\}}}\psi_{j_{\{O\}} j_{\{H\}} j_{\{I\}}}|j_{\{O\}}> |j_H >|j_{\{I\}}>
\ee
The index for the horizon edge is written separately as they occupy a special status, as they
are more or less free wavefunctions which are factorisable from the rest of the wavefunction. $j_{\{O\}}$
denotes all the spins labelling the outside edges and $j_{\{I\}}$ denotes all the spin labels for the
inside edges.
The horizon wavefunction however is determined by the area $P_H$ for the horizon bit that the
edge induces. Infact, it is precisely the $P_H$ of the horizon edge which can be
considered as a variable and adds to the entropy of the horizon. The form of the coherent state
wavefunction remains the same as above, except that now the labels $g_{v_2}$ of the wavefunctions 
inside the horizon, are functions of those outside. The exact functional form will depend on the solution from the apparent horizon equation.{\it Moreover, even so, the correlation is difficult to quantify in a entropy calculation. To enable the entropy calculation to first
order in the semi-classical parameter, we encode the correlation in a conditional probability function
$f(g_{r_o},g_{\th_o}|g_{r_i},g_{\th_i})= N $,
(N is a constant), when the values are as determined by the apparent horizon equation or is 0 otherwise.} 
Given the formalism discussed in the paper, this is the only way one can introduce a correlation in the
wave function. The `classical correlation' is encoded in the expectation value of the operators,
and hence the labels $g_e$. The functional form of the wavefunction and it's dependence on $g_e$
remains unchanged, as obtained from the coherent state transform. 

\includegraphics[scale=0.8]{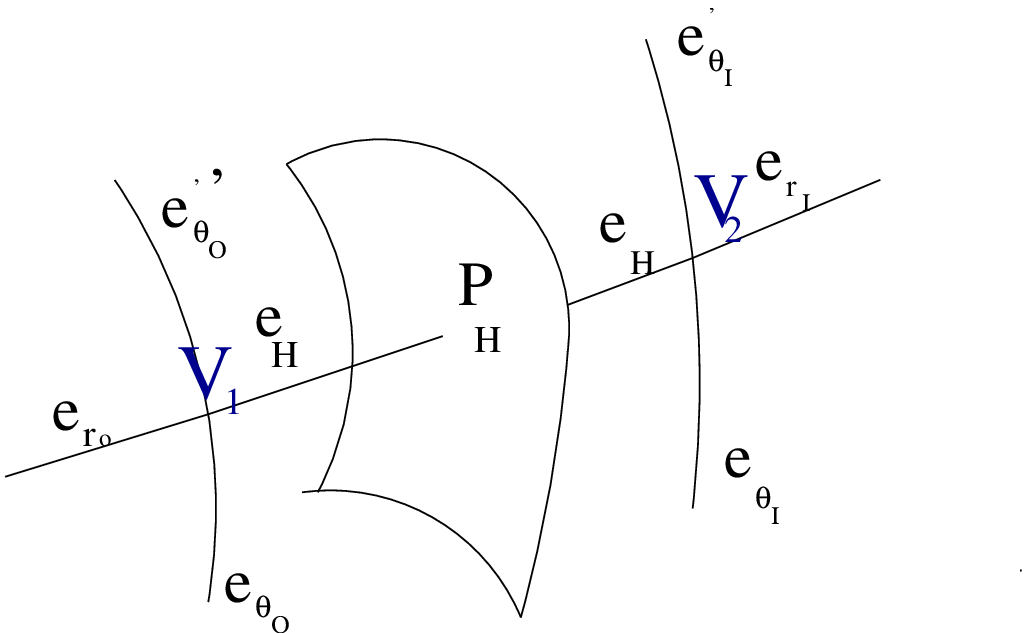}

\be
|\Psi(v_1,v_2)>= \sum_{j_{\{O\}} j_H j_{{\{I\}}}}\psi_{j_{\{0\}} j_H j_{\{I\}}}(\pi,f) |j_{\{O\}}>|j_H>|j_{\{I\}}>
\ee
Here, the coefficient $\psi_{j_{\{O\}} j_H j_{\{I\}}}$ now includes the correlation function f. 
Now clearly the above coherent state is written for the edges which are connected across the horizon
by one horizon edge. In otherwords they share the vertex $v_1$ and $v_2$ which are the initial
point and final point of the radial edge crossing the horizon. Thus the density matrix constructed
of the above state can now be written in the form:
\be
\rho_{v_1, v_2}=|\Psi(v_1,v_2)><\Psi(v_1,v_2)|
\ee
The full density matrix for the entire horizon would comprise of
\be
\rho=\prod_{v_1,v_2} |\Psi(v_1,v_2)><\Psi(v_1,v_2)|
\ee
Where the product is over all the vertices comprising the immediate outside and inside
of the horizon. The tracing is now done over the edges which share $v_2$, or the internal
edges. The reduced
density matrix then has the following components (following the same procedure as in Equation(\ref{eq:trace})):
\be
\rho_{v_1 \ j_{\{O\}} j_H,j'_Hj_{\{O'\}}}= \sum_{\{I\}}\bar{\psi}_{j_{\{O\}} j_H j_{\{I\}}}\psi_{j_{\{O'\}} j'_H j_{\{I\}}}
\ee
Explicitly in terms of the coherent state wavefunction, the sum results in terms of the form:
\bea
\rho_{v_1 \ j_{\{O\}} j_H, j'_H j_{\{O'\}}} & = & \sum_{j_{\{I\}}}\frac{ \bar\pi_{j_{\{O\}}}}{||\psi||^{1/2}}\frac{\bar\pi_{j_H}}{||\psi||^{1/2}}\frac{\pi_{j_H'}}{||\psi||^{1/2}}\frac{\pi_{j_{\{O'\}}}}{||\psi||^{1/2}}\frac{|\pi_{j_{{\{I\}}}}(g)|^2}{||\psi||} |f_{j_{\{O\}},j_{\{I\}}}|^2 e^{-t j_{\{I\}}(j_{\{I\}}+ 1)}  \\  &\times &  e^{-tj_{\{0\}}(j_{\{0\}}+1)/2}e^{-tj_H(j_H +1)/2}e^{-t j'_H (j'_H +1)/2} e^{-t j'_{\{O\}}(j'_{\{0\}} +1)/2}
\eea
The result of the summation of the indices $j_{\{I\}}$ is a Gaussian, peaked at the classical value
of $g_I$. 
(Note $j_{\rm cl}$ for any of the edges outside, inside, or at the horizon indicates a peak
value which equates to classical parameters.)

The density matrix which is reduced now has the following form
\be
\rho_{v_1 \ j_{\{O\}} j_H j'_{\{O'\}}} = \frac{\bar\pi_{j'_{\{O\}}}\pi_{j_{\{O\}}}}{||\psi||} e^{-t j_{\{O\}}(j_{\{O\}} + 1)/2}e^{-t j'_{\{O\}}(j'_{\{O\}} +1)/2} \frac{\bar\pi_{j_H}\pi_{j_H}}{||\psi||} e^{-t j_H(j_H+1)} |f_{j_{\{O\}},j_{\{ I \rm cl\}}}|^2 \ee Due to the $||\psi||$ in the denominator which contributes with a factor proportional to $e^{-P^2/t} e^{-t/4} \sinh(p)/p$, the

(Note ${cl}$ for any of suffixes for edge labels outside, inside, or at the horizon indicates a peak
value which equates to classical parameters.)

offdiagonal elements are all damped exponentially in the semi-classical regime where $t\rightarrow 0$,
and $P$ is very large. For the diagonal elements this factor is assimilated into a Gaussian which
becomes a delta function in the semiclassical limit. Infact as determined in \cite{tow}, this implies
\be
\frac{\bar\pi_{j} \pi_j}{||\psi||}e^{-t j(j +1)} \sim e^{-\frac j2\frac{(m/j - ^R P_3/P)^2}{(1-^R P_3/P)^2}} e^{-\frac j2\frac{(n/j - ^LP_3/P)^2}{(1-^RP_3/P)^2}} e^{- \frac{\left[(j+1/2)t -P\right]^2}{t}} 
\label{eq:prob}
\ee
The interesting aspect of this is that when the semiclassical parameter $t\rightarrow 0$, the 
functions in (\ref{eq:prob}) tend to delta functions, each peaked at appropriate values of the
classical variables.
Thus, the `density matrix' reduces to a set of delta functions:
\be
\rho_{v_1 {j_{\{O\}}} j_{H},{j_{\{O\}} j_H} }= \delta(j_{\{O\}}t, P_{\{O\}})\delta(m_{\{O\}}t, P^I_{\{O\}}) \delta(j_H t, P_H)\delta(m_H t P^I_H) |f_{j_{\{O\}},j_{ I \rm cl }}|^2
\label{eq:den}
\ee
The density matrix has non-zero values
for only a set of matrix elements, as determined by the delta function equations. 
 The non-zero elements are 
$\rho_{v_1 j_{\{O\} {\rm cl}} j_{H {cl}} m_{H {\rm cl}}, j_{H {\rm cl}} m_{H {\rm{cl}}} j_{\{O\} \rm cl}}$, however, the 
$m_{H \rm cl} =- j_{H \rm cl} ..... j_{H \rm cl}$. At the horizon only $P_H$ is
fixed, and hence the entire range of $m_H$ is allowed.
This is the origin of the degeneracy of the density matrix, and the entropy of the entire
space-time. After all the discussion on the coherent states and the density matrices, the
entropy is finally the number of ways to build the horizon area. Inclusion of all the edges
comprising the entire graph in the initial density matrix would not contribute to the entropy,
as they would belong to tensor product of Hilbert spaces, and give unit norms in the trace. 
However, one must emphasise
that this is a `semi-classical' result, valid at 0th order in the semi-classical parameter
$t$. If one includes the higher order corrections, one would get a `quantum entropy' and the
importance of the above result for an explicit expression for the density matrix is in the
obtaining of the corrections to black hole entropy. 
From (\ref{eq:den}), and $Tr(\rho)=1$, it follows that $|f_{j_{O\rm cl},j_{I \rm cl}}|^2 = \frac{1}{2 {j_{H {\rm cl}}} +1}$.
This apparently `trivial' factor $f$ still has non-trivial information about the classical correlations, as now
$j_{O \rm cl}$ and $j_{I \rm cl}$ are determined by the classical labels of the edges at the vertex $v_1$ and 
$v_2$. Hence, if the $j_{O \rm cl}$ and $j_{I \rm cl}$ were anything other than those determined by the apparent horizon
equation, then $f$ would be zero.
Now the full density matrix will be a product of all the density matrices of the vertices immediately
outside the horizon. Thus the Entropy would be:
\be
S_{BH}= -Tr[\rho\ln \rho]= -Tr[\prod_v \rho_v \ln (\prod_v \rho_v)]= \sum_v \ln(2 j_v +1)
\ee

The additional constraint is
\be
\sum\left(j_v +\frac12\right)t = \frac{A_H}{a} \ \ {\rm or} \sum\left(j_v + \frac12\right)=\frac{A_H}{l_p^2}
\ee
In case all the area bits are set as equal $j_v= j_s$, one obtains, 
\be
N(j_s + \frac12)= \frac{A_H}{l_p^2}
\ee

and Entropy
\be
S_{BH}= \frac{A_H}{(j_s + 1/2) l_p^2}\ln ( 2 j_s + 1)
\ee

This obviously differs from the Bekenstein Hawking entropy due to the dependence on Spin.
We will subsequently discuss the Immirzi parameter which will be adjusted to give the
appropriate entropy. 

\subsection{Sum over all possible graphs}
In the previous discussion, the entropy is shown to arise due to the degeneracy associated with
the number of ways of inducing horizon area, given a graph. The graph is such that only
radial edges cross the given horizon. Now, this assumption can be somewhat generalised to
include a sum over all possible similar graphs, which would imply a difference in the number
of edges crossing the horizon. This then would include a sum over all possible areas induced by
a horizon edge, though the entire formalism of the previous section would remain absolutely same. The coherent state formalism is useful is studying this sum over all possible $P_H$ for the reason, that, there exists a Liouville measure in the classical phase space, with respect to which the states are overcomplete. 
Now, when the density matrix for a particular set of vertices is defined as
\be
\rho_{v_1 v_2}= |\Psi(v_1,v_2)><\Psi(v_1,v_2)|
\ee
The apparent horizon equation fixes the value of the correlated 
wavefunction $\psi_{j_{\{O\}}j_H j_{\{I\}}}$ at a particular value, as all
the classical labels are determined. However, if now one chooses to integrate
over all possible areas induced at the horizon, then, the classical labels
for the $g_{\th_o}$ and $g_{\th_i}$ are also going to change accordingly. To
quantify the above, one can thus write the conditional probability as
now a function of $P_H$, with a $f(g_{\th_o},g_{\th_i}|P_H)=N$, if the
classical labels are as obtained from the apparent horizon equation, or
zero otherwise. Once the density matrix is obtained, then the diagonal terms
give the probability of finding the system in that state, given any internal state. Now, this
probability function is multiplied by the `overlap function' or the probability $p^t(g_H,g'_H)$,
with respect to the Liouville measure. The probability function gives the probability of finding
the system at the phase space point $g'$, when it is in a state $\psi^t(g_H)$. 

\be
p^t(g_H,g'_H)d\Omega = p^t(g_H,g'_H)e^{-P_H^2/t} e^{-t/4} \frac{2\sqrt 2}{(2\pi t)^{3/2}}\frac{\sinh P_H}{P_H} d^3 P_H d\mu_H
\ee

Where $d\mu_H$ is the Haar measure. Now, finally, the probability and the correlation function
are independent of the holonomy, and so, the Haar measure integrates out to 1. Also, in the
$t\rightarrow0$ limit, the ovelap function is almost 1. Thus, one retaines now only the Liouville
measure in the variable $P_H$.  
Now the measure $d^3P_H$ can be converted to $P^2_H dP_H \sin\Theta d\Theta d\Phi$, where the angles
$\Theta$ and $\Phi$ are in the internal 3 space. Integration over the angles would achieve 
a averaging over the components of the momentum, with it's length as fixed.

The diagonal terms in the density matrix dependent on the horizon wavefunction are of the form:

\be
= \left(\pi_j(H^2)_{mm} e^{-tj(j+1)}\right)e^{-P_H^2/t} e^{-t/4}\frac{d\Theta d\Phi}{t^3/2}\sin\Theta  P_H^2 dP_H
\ee
The expression for $\pi_j(H^2)_{mm}$ is then the following:
\bea
|\pi_j(H^2)_{mm}|^2 &= & \sum_l \frac{(j+m)!(j-m)!}{(j-m-l)!(j+m+l)!(l!)^2}\left[\cosh^2 P_H - \cos^2\Theta\sinh^2P_H\right]^j \nn \\
&\times &\left[\frac{\cosh P_H + \cos\Theta\sinh P_H}{\cosh P_H -\cos\Theta\sinh P_H}\right]^m\left[\frac{\sin^2\Theta\sinh^2 P_H}{\cosh^2P_H -\cos^2\Theta\sinh^2P_H}\right]^l
\eea
Using this back in the equation, the above leads to a integrand which can be reduced to:
{\small \be
\sum_l\sum_k\sum_{k'}M_{jlm} \cosh^{2j} P_H \left[\tanh P_H\right]^{2 (j - k -k'/2)}(-1)^{j -m -l -k}  C^{j-m-l}_k  C^{2m}_k \left[\cos\Theta\right]^{2(j-m-l-k)}\sin^{2l}\Theta\cos^{2m-k'}\Theta \nn
\ee}
with
\be
M_{jlm}= \frac{(j+m)!(j-m)!}{(j-m-l)!(j+m+l)! (l!)^2}
\ee
The integral over $\Theta$ is $\int_0^{\pi} \sin^{2l +1}\Theta \cos^{2j-2l -2k -k'}\Theta d\Theta$ and hence yields a beta
function $B(2l+2, 2j - 2l -2k -k'+1)$, with the additional restriction that $k'=2n$, such that the
sum is non-zero, the sum now reduces to 
\be
\sum_l\sum_k\sum_n M_{jlm}\cosh ^{2(j)} P_H \left[\tanh P_H\right]^{2(j -k-n)} (-1)^{j-m-l-k} C^{j-m-l}_k C^{2m}_{2 n} B(2 l +2, 2j - 2l -2k -2n +1)
\ee
with
\be
M_{jlm}= \frac{(j+m)!(j-m)!}{(j-m-l)!(j+m+l)! (l!)^2}
\ee
Defining $k+n = q$ one gets:
\be
\sum_l\sum_k\sum_n M_{jlm}\cosh ^{2(j)} P_H \left[\tanh P_H\right]^{2(j-q)} (-1)^{j-m-l-k} C^{j-m-l}_k C^{2m}_{q-k}\frac{\Gamma(2l+2)\Gamma(2j -2l -2q+1)}{\Gamma(2l-2q +3)}
\ee
This is thus the expression for the density matrix element, and a integral over the Liouville measure will give the `sum over all possible graphs' entropy. 

There is a interesting observation due to the $e^{-P_H^2}/t$ term in the Liouville measure, this is discussed here.  The calculation simplifies 
due to the observation, that classical $P_H= \sin\th_0\th\phi$, if one takes $a=r_g^2$, the only possible 
area scale in the system, and hence independent of the
black hole horizon radius! The value of $P_H$ should vary from 0 to $\infty$, depending on the number of
edges crossing the horizon (increase in number of edges decreases the value of $P_H$). Also, due to the presence of the $e^{-P^2_H/t}$ term in the measure, the
maximum value of the density matrix elements are concentrated around $P_H=0$. In the regime where $P_H\approx 0$, the $l=0$ term in the sum dominates,
and the sum trivially reduces to 1. This concentration of the possible Horizon area elements near a $(j_{cl}+1/2)t\approx 0$ value
is in agreement with previous derivation of black hole entropy as arising from $j=1/2$ spin elements, albeit
in different formalisms \cite{asht}. The density matrix elements then are of the form:
\be
\rho_{jj} = |f|^2
\ee
Setting $\sum_m|f|^2=1$, one obtaines $|f|^2 = 1/(2j_{cl}+1)$, and hence the correct value of entropy
is recovered. This limit is in the situation where the number of edges crossing the horizon are
large, and quite opposite to the regime discussed in the previous fixed graph calculation. However, in both the regimes the same entropy law, proportional to
area of the horizon is recovered. Further, in this regime, the offdiagonal elements of the density
matrix will also be non-zero, and provide corrections terms to the entropy.
 
\section{Entropy and Immirzi parameter}
This factor prevents the above entropy counting from being the exact Bekenstein-Hawking entropy.
However, what has been excluded from the above is the fact that the area spectrum is $(j_{cl}+1/2)\beta t$,
for different quantisation sectors of the theory. Since the constraint is $N(j_s+\frac12) l_p^2=A_H$, where
$A_H$ is the classical area of horizon, which remains unchanged for the Immirzi parameter related 
variables. 
\be
N=\frac{A_H}{\beta (j_s + 1/2)}\ln(2 j_s + 1)
\ee

Thus it seems that if $\beta= \frac{4 (j_s + 1/2)}{\ln (2 j_s + 1)}$ would be the correct choice for the
Immirzi parameter. However, this is rather dependent on the choice of graph, and the area it induces on the
horizon. In the most general situation, this would include a sum over all possible graphs, and the coefficient
of the counting would equate the value of the Immirzi parameter. Here the counting would be different from that
given in \cite{meiss} and hence the value of the Immirzi parameter would be considerably different, from the
prediction there. 

\section{Entropy of Space-times with Apparent Horizons}
Here we summarise the results in the previous sections for the entropy of the Schwarzschild black hole, and show
how they can be easily extended to include spherically symmetric apparent horizons.
The key assumption of the entire article has been that the classical spatial slice, at which
the solution is peaked has spherical symmetry. In otherwords, the metric on the
spatial slice has the form
\be
ds^2= dr^2 + r^2 d\Omega
\ee
This also presumes that the intrinsic metric is flat. The information about the curvature of the
entire slice is contained in $K_{ab}$ or the extrinsic curvature for the entire slice. The
presence of the apparent horizon equation is obtained as a solution of the equation
\be
-\G^{\th}_{\th r} - \G^{\phi}_{\phi r} + K_{\th}^I e^{\th}_I + K_{\phi}^I e^{\phi}_I=0
\ee
Where $K_{\th \th}$, $K_{\phi \phi}$ are arbitrary functions of the radial coordinate.
As before the graph is taken along the radial, and angular coordinate lines, forming
discrete lines. The classical space is thus sampled in terms of discrete variables
$h_e, P^I_e$, which are then combined to form the classical label on which the
coherent states are peaked. Since the apparent horizon equation places no conditions on the
radial edges, a similar construction is done, where horizon areas are induced only by the
radial edges. The apparent horizon equation then is a difference equation linking the
vertices which the radial edges connect. The coherent states are then correlated across the vertices.
The density matrix and the subsequent derivation of entropy thus follows automatically. This 
would imply e.g., any spherically symmetric apparent horizon will have entropy proportional
to the area of the horizon, this would then automatically include cosmological horizons, also.
The addition of a cosmological term in the lagrangian does not change the definition of the kinematical
variables of the theory, and hence all the discussions in the previous sections of the coherent states does not change the 
derivation of entropy. 
\section{Conclusion}
In this paper, the main aspects of black hole
physics emphasised are:

\noindent
1)Black holes are classical solutions of Einstein's equations and hence would correspond to a 
semi-classical state like the coherent state in a quantum theory.\\
2)In a exact quantisation of the theory, it is difficult to identify a complete `black hole'
state, though, states with trapped surfaces in reduced phase space quantisations have been
obtained in \cite{huwi}, and also the isolated horizon boundary condition basically identifies
the horizon by imposing the boundary condition on the physical states of the theory.\\
3) The coherent states defined in \cite{thiem} on the other hand provide a complete
formalism for the state which would correspond to a given black hole solution.\\
4)The given coherent state which is peaked in both the momentum and configuration 
space variable is also a state where both the `area' (intrinsic metric) and the extrinsic
curvature (holonomy) can be measured with minimum uncertainty.\\
5)Thus the measurements in the coherent state allows one to build a entire spatial slice
of the given space-time, and hence locate the apparent horizon and the singularity uniquely.\\
6)The information about the space-time is however encoded in regularised variables of the
holonomy defined along one dimensional edges of finite length, and corresponding momentum 
induced on dual surfaces of the edges. This regularisation in case of a singularity at one
point effectively regularises the same. The holonomy which has information about the curvature
never diverges and takes oscillatory values from -1..1. \\
7)A clear uncertainty principle prevents measurement of area zero, and hence, the singularity
in the curvature operator is resolved mainly due to the existence of a minimal area.\\
8)The Coherent state is a tensor product of coherent states over edges, and hence one can
trace over the edges inside the horizon.\\
9)To encode the correlation of the classical variables across the horizon, one finds that the
apparent horizon equation is a function of the variables outside the horizon, as well as those
inside the horizon.\\
10)The wavefunctions which are functions of the classical labels are thus correlated, and 
lead to a entropy of the resultant density matrix obtained after tracing the wavefunction inside
the horizon. It shall be very interesting to calculate the corrections to entropy, and whether they
have a logarithmic behaviour as calculated in \cite{corr}. \\
There are quite a few questions still to be completely understood.\\
1) What exactly happens for a general spatial slicing, i.e. those without the flatness assumption?\\
2)Are there more coherent states than the ones stated here, and what would the expectation values
of operators in those states be?\\
For immediate future problems:\\
1)The above formalism can be generalised to arbitrary black holes, those with charge and angular momentum.\\
2)The regularised curvature operator in terms of holonomy and momenta can be measured in other
states, not necessarily semiclassical.\\
3)The semiclassical derivations of black hole thermodynamics here can be used to obtain a approximate
temperature using the density matrix, and a evolution in time. This is work in progress.\\
To summarise, the semi-classical nature of black holes have been discussed. The quantisation, is non-perturbative,
and includes the full degrees of freedom for gravity. It will be interesting to verify the Hamiltonian
constraint completely for the coherent states, right now the action of the Hamiltonian on these states leads 
to states of very small norm. There is a discretisation of the Schwarzschild space-time using dynamical triangulation techniques
in \cite{bianca}, it shall be interesting to obtain the entropy in that formalism.

{\bf Acknowledgements}. I would like to thank S. Das, B. Dittrich, J. Gegenberg, G. Gour, V. Husain, V. Suneeta, T. Thiemann
for useful discussions during various stages of this work. Part of the work was started at ULB, 
and University of Lethbridge. This work is supported by the research funds of University of New Brunswick.

\appendix{Appendix A: The Quantum Action Principle} \label{app:qac}\\

The Quantum action principle is postulated as
\be
S_{Q} = \int d \t <\xi| \frac{d}{d\t}|\xi>
\ee
Applying it to the restricted set of Coherent states, one gets the following expression:
\be
S_{Q res} = \int d\t [\int \frac{\bar{\psi}^t_g(h)}{||\psi||} \frac{d}{d\t} \frac{\psi^t_g(h)}{||\psi||} d\mu(h)]
\ee
($d \mu(h)$ corresponds to the Haar measure) Now, to quote \cite{tow}, the state
\be
\psi^t_g(h) = \sum_{j} (2j+1) e^{-t/2j (j+1)}\chi_j(g h^{-1})
\ee
As calculated,
\be
\chi_j(gh^{-1}) = \frac{\lambda^{2j+1} - \lambda^{-(2j+1)}}{\lambda - \lambda^{-1}}
\ee
with
\be
\lambda = x + \sqrt{x^2 -1}
\ee
where $x= \cosh(P/2)\cos\th + \imath \frac{(P^j\th^j)}{P\th}\sinh(P/2)\sin(\th)$.
Thus
\be
\frac{d}{d\t} \chi_j(gh^{-1}) = -\frac{\dot\lambda}{\lambda}\left(\chi_j(gh^{-1})\frac{\left(\lambda + \lambda^{-1}\right)}{\left(\lambda - \lambda^{-1}\right)} + (2j+1)\frac{\l^{(2j+1)}+ \l^{-(2j+1)}}{\l -\l^{-1}}\right)
\ee

Putting this back into the scalar product, one obtaines:

\bea
<\psi|\frac{d}{d\t}|\psi> &= & \int d\mu(h) \frac{\dot\lambda}{\lambda}\left[\left( \frac{\left(\lambda + \lambda^{-1}\right)}{\left(\lambda - \lambda^{-1}\right)}\right)\bar\psi\psi + \bar\psi\left(\sum_j(2j+1)(2j+1)e^{-tj(j+1)/2}\frac{\l^{2j+1} + \l^{-(2j+1)}}{\l - \l^{-1}}\right)\right] \nn \\& - & \frac1{||\psi||}\frac{d}{d\t}||\psi||
\label{eq:abc}
\eea
The first term in the above integrand clearly is proportional to the probability density, which as $t\rightarrow 0$,
is peaked at the value of $\th=0$. We investigate the second term there, to see if the sum can be converted
to a convergent expression. The technique used there are as in \cite{tow}. We concentrate on the sum first:
\bea
&& \sum_j (2j+1)^2 e^{-tj(j+1)/2}\frac{\lambda^{(2j+1)} + \lambda^{-(2j+1)}}{\lambda -\lambda^{-1}} \nn \\
&=& \frac1{\lambda -\lambda^{-1}}\sum_j (2j +1)^2 e^{-(t/8)(2j(2j+1))}\left[e^{(2j+1)\ln\lambda} - e^{-(2j+1)\ln\lambda}\right] \nn \\
&=& \frac1{\lambda -\lambda^{-1}}\sum_{n=-\infty}^{\infty} n^2 e^{-t/8[n^2 - 8 n\ln\lambda/t]} \nn \\
&=& e^{t/8} \frac1{\lambda -\lambda^{-1}}\sum_{n=-\infty}^{\infty} n^2 e^{-[(ns)^2 - 2(ns)z]/2}
\eea
with $n=2j+1$ and $s^2=t/4$, $z= \frac{\ln\lambda}{s}$.

This is converted by the Poisson Summation formula to a convergent series by using the following property
\be
\sum_{n=-\infty}^{\infty} f(ns) = \frac{2\pi}{s}\sum_{n=-\infty}^{\infty} \tilde {f} \left(\frac{2\pi n}{s}\right)
\ee

where $\tilde f$ is the fourier transform of $f(ns)$.

Here,

\be
s^2 f(ns) = (ns)^2 e^{-[(ns)^2 -(ns)z]/2}
\ee

The fourier transform of which can be easily evaluated as the integrals are simple Gaussians.

\be
\tilde f(k)= \frac1{\sqrt{2\pi}}\left[2 + (z-ik)^2\right]e^{\frac12(z-ik)^2}
\ee

Substituting the same in the Poisson summation formula, one obtaines

\be
\sum_{n=-\infty}^{\infty}f(ns)= \frac{\sqrt{2\pi}}{s^3}\sum_{n=-\infty}^{\infty}\left[2 + \left(z- i \frac{2\pi}{n}\right)^2\right]e^{\frac12\left(z -i \frac{2\pi}{n}\right)^2}
\ee
As the sum is computed, in the $t\rightarrow 0$ limit, only the following terms are relevant
\be
\approx \frac{\sqrt{2\pi}}{s^3} e^{t/8}\frac{1}{\l - \l^{-1}}\frac{[\cosh^{-1}(x)]^2}{s^2} e^{\frac1{2 s^2}[\cosh^{-1}(x)]^2}
\approx \frac{[\cosh^{-1}(x)]}{s^2}\psi
\ee
with $\ln\lambda=\cosh^{-1}(x)$.
Substituting in the expression for the action, this term in the action, one obtaines the following set
of terms:
\be
I_{resquantum}= \int d\mu (h)\frac{\dot\lambda}{\lambda}\left[\frac{(\l +\l^{-1})}{\l-\l^{-1}} + \frac{\cosh^{-1}(x)}{s^2}\right]p^t(h)
\ee
where $p^t(h)= \bar\psi \psi/||\psi||^2$. 
Now,
\be
\frac{\dot\lambda}{\lambda} = \frac{\dot x}{\lambda- \lambda^{-1}}
\ee

with

\bea
\dot x &= &\frac{\dot P}{2} \sinh\left(\frac{P}{2}\right)\cos\th + \sin\th \dot\th \cosh\left(\frac{P}{2}\right) \nn \\
&+& \imath\frac{\dot P^j\th^j  + P^j\dot \th^j}{P\th} \sinh\left(\frac{P}{2}\right)\sin\th  -\imath\frac{P^j\th^j}{P^2\th^2}\left(\dot P \th + \dot \th P\right) \sinh\left(\frac{P}{2}\right) \sin\th
 \nn \\
&+& \imath\frac{P^j\th^j}{P\th}\cosh\left(\frac{P}{2}\right)\dot P \sin\th + \imath\frac{P^i\th^i}{P\th} \sinh\left(\frac{P}{2}\right) \cos\th \dot \th
\eea

Substituting the above in equation (\ref{eq:abc}), one obtains the terms which survive after the
integration of inner product, (one replaces the terms with $\th,\th^j$ with 0, as the coherent state
is peaked precisely at these values),
\be
\int \frac{\bar\psi}{||\psi||^2} \frac{d}{d\t} \psi= \frac1t\left[P\frac{\dot P}{2} + \imath \dot\th^jP^j\right]
\label{eq:abcd}
\ee
also,
\be
||\psi||^2= \sum_j (2j+1) e^{-tj(j+1)} \chi_j(H^2)
\ee
one obtains:
\be
\frac{d}{d\t}||\psi||^2 = \frac{\dot{\xi}}{\xi}\sum_j(2j+1)\left[(2j+1)\left(\frac{\xi^{2j+1} + \xi^{-(2j+1)}}{\xi-\xi^{-1}}\right) - \frac{\xi + \xi^{-1}}{\xi - \xi^{-1}}\chi_j(H^2)\right]e^{-t j(j+1)}
\ee
Given the fact that $\xi= \cosh P + \sinh P$, the above can be calculated, and in the $P\rightarrow\infty$ limit, it gives:
\bea
\frac{1}{2 ||\psi||^2}\frac{d}{d\t} ||\psi||^2 & = & \frac12\frac{\dot \xi}{\xi ||\psi||^2}\sum_j (2j+1)^2\left\{\frac{\xi^{2j+1} + \xi^{-(2j+1)}}{\xi -\xi^{-1}}\right\}e^{-t j(j+1)} \\&& -\frac12\frac{\xi + \xi^{-1}}{\xi -\xi^{-1}}\frac{\dot\xi}{\xi}
\eea
The second term in the above yields in the $P\gg 1$ limit
\be
-\frac12 \dot P\coth P\approx -\frac12 \dot P
\ee
The first term needs a Poisson resummation formula implemented (z=P):
\bea
&& \frac1{\xi - \xi^{-1}}\sum_j (2j+1)^2 \left\{\xi^{2j+1} + \xi^{-(2j+1)}\right\}e^{-t j(j+1)}\\
& = & \frac{8 \sqrt{\pi}}{t^{3/2}\sinh p}\sum_{k=-\infty}^{\infty} e^{(z-ik)^2/t}\left[1 + \frac{(z-ik)^2}{t}\right]\\
&\approx& \frac{8 \sqrt\pi}{t^{3/2}} e^{P^2/t}\left(1 + \frac{P^2}{t}\right)
\eea
Thus finally, in the limit $P\rightarrow \infty$,
\be
\frac1{2 ||\psi||^2}\frac{d}{d \t}||\psi||^2 = \frac12\dot P\left(\frac1{P} + \frac{P}{t}\right) - \frac12\dot P
\ee

Interestingly, substituting the above in equation(\ref{eq:abc}), one obtaines the cancellation of the
first term in Equation(\ref{eq:abcd}), and finally one obtaines the expression for the classical action.

\appendix{Appendix B: Trace terms in the Curvature Operator}\label{traceterms}\\

Since the trace terms take rather long expressions for certain cases, they are described in the Appendix here,
though the physical interpretation for them is exactly same for each of the terms,
and the conclusion is that the expectation value of the curvature operator is bounded. In the evaluation of the
terms which are of the form
\be
<\psi|{\rm Tr}[h_{e_a}^{-1}[h_{e_a},V]h_{e_b}]|\psi>
\ee
The following traces occur:
\bea
{\rm Tr} (g^{\dag}_a g_a g_b)e^{-9t/8} &= &2 \cos\s_b\cosh P^a \cosh\frac{P^b}{2} + 2 \cos\s_b \frac{P_I^a P_J^b}{P^a P^b}\sinh{P^a}\sinh{\frac{P^b}{2}} \nn \\
&& - 2 i \s_b^I P_b^I \frac{\sin{\s_b}}{\s_b} \frac{\sinh P^b/2}{P^b}\cosh{P^a} - 2 i \s^b_I P^a_I \frac{\sin{\s_b}}{\s_b}\frac{\sinh P^a/2}{P^a}\cosh (P^b/2) \nn \\
&& -2\e^{IJL}P^a_IP^b_J\s^L_b \frac{\sin\s_b \sinh{P^b/2}\sinh P^a}{\s_bP^a P^b} -2i P^a_I\s_a^J\frac{\sinh P^a \sin\s_a}{P^a \s_a}\e^{IJK} \nn\\
&\times&\left[2\s^k_a\cos\s_b\cosh(P^b/2)\frac{\sin\s_a}{\s_a} - 2 i \e^{KMN}\s_a^MP^B_N\cos\s_b\frac{\sin\s_a\sinh P^b/2}{\s_a P^b} \right.\nn\\
&& - 2 \s_b^L\cos\s_a\cosh(P^b/2)\frac{\sin\s_b}{\s_b} - 2 i \e^{KMN}\s^M_b P_b^N \cos\s_a\frac{\sin\s_b\sinh P^b/2}{\s_b P^b} \nn\\
&& +\left. \e^{MNK}\s_b^M\s_a^N \cosh(P^b/2)\frac{\sin\s_b\sin\s_a}{\s_b\s_a} \right]
\eea
The other trace term is 
\be
{\rm Tr}\left[g^{\dag}_aT^I g_a g_b \right] e^{-9t/8}  = {\rm Tr}\left[(e^{-i P_a T/2} h_a)(e^{-i P_b T/2} h_b)(h_a^{-1} e^{-i P_aT/2}) T^I\right]\nn   \\
\ee
which can be simplified to
\bea
&& {\rm Tr}\left[e^{-i P_a T/2}h_a (h_a^{-1}) e^{-i P_bT/2} h_b e^{-i P_aT/2} T^I\right] + {\rm Tr}\left[(e^{-i P_a T/2} h_a [g_b,h_a]e^{-i P_a T/2} T^I\right] \nn \\
&=& {\rm Tr}\left[h_b e^{-i P_aT/2} T^I e^{-i P_a T/2} e^{-i P_b T/2}\right] + 2 {\rm Tr}\left[h_a e^{-i P_b T/2} T^K e^{- i P_a T/2}\right]\e^{IJK}\s_b^I\s_b^J\frac{\sin\s_a}{\s_a}\frac{\sin\s_b}{\s_b} \nn \\
 &-2 i& {\rm Tr}\left[h_a T^k h_b e^{- i P_a T/2}\right] 
= \cos\s_b {\rm Tr}\left[e^{-i P_a T/2} T^I e^{-i P_a T/2} e^{- i P_b T/2}\right] \nn \\ &+ & \frac{\s_a^J\sin\s_a}{\s_a}{\rm Tr}\left[T^J e^{-i P_a T/2} T^I e^{-i P_a T/2}e^{-i P_b T/2}\right] \nn\\
&=& 2 \e^{IJK}\s_b^I\s_a^J\frac{\sin\s_a}{\s_a}\frac{\sin\s_b}{\s_b} {\rm Tr}\{\cos\s_a\left[e^{-i P_bT/2}T^K e^{- i P_aT/2}\right] \nn \\
&+& \frac{\s_a^L\sin\s}{\s}\left[T^L e^{-i P_b T/2}T^K e^{-i P_a T/2}\right] \} \nn \\
& -& 2i \left[\cos\s_a\cos\s_b - \s_a\s_b \frac{\sin\s_a}{\s_a}\frac{\sin\s_b}{\s_b}\right]{\rm Tr}[T^K e^{- i P_aT/2}] \nn\\
&+& \left[\e^{IJL}\s^I_a\s^J_b + \s^L_a \cos\s_b \frac{\sin\s_a}{\s_a} + \cos\s_a\s_b^L\frac{\sin\s_b}{\s_b}\right]{\rm Tr}\left[T^L T^K e^{- i P_a T/2}\right] \nn \\
& + &2 \cos\s_a {\rm Tr}\left[ T^L e^{-i P_a T/2}\right] + \frac{\s^K\sin\s}{\s}{\rm Tr}\left[T^K T^L e^{-i P_a.T/2}\right]
\eea

The remaining expression in the traces here are all bounded terms and hence it is sufficient to show
from the above that none of the terms in the curvature operator diverge.

One also evaluates the following then using exactly similar techniques, $ {\rm Tr}\left[g^{\dag}_a g_a T^I g_b\right] {\rm Tr}\left[g_b g^{\dag}g_a T^I\right]$.

\end{document}